\documentclass[a4paper,10pt]{article}
\usepackage{import}
\usepackage[T1]{fontenc}
\usepackage[utf8]{inputenc}
\usepackage[english]{babel}

\usepackage{amsmath, nccmath}
\usepackage{amssymb}
\usepackage{amsthm}
\usepackage{enumerate}
\usepackage{enumitem} 
\setlist[enumerate]{wide = 0pt, leftmargin=*}
\usepackage{appendix}
\usepackage{geometry}
\usepackage{eurosym}
\usepackage[hyphens]{url}
\usepackage{hyperref}
\hypersetup{
	colorlinks=true,
	citecolor=blue,
	filecolor=blue,
	linkcolor=blue,
	urlcolor=blue
}
\usepackage[hyphenbreaks]{breakurl}
\usepackage[sort&compress]{cleveref}
\usepackage{todonotes}
\usepackage{mathrsfs}

\usepackage{rotating}
\usepackage[round]{natbib}

\def\be{\begin{eqnarray}}
\def\ee{\end{eqnarray}}

\def\b*{\begin{eqnarray*}}
	\def\e*{\end{eqnarray*}}

\newtheorem{Definition}{Definition}[part]
\newtheorem{Proposition}{Proposition}[part]

\newtheorem{AssumptionAppendix}{Assumption}
\newtheorem{Lemma}{Lemma}[part]
\newtheorem{Corollary}{Corollary}[part]
\newtheorem{Remark}{Remark}[part]
\newtheorem{Example}{Example}[part]

\newcommand{\ba}{\begin{array}}
	\newcommand{\ea}{\end{array}}
\newcommand{\ben}{\begin{equation*}} 
\newcommand{\een}{\end{equation*}}
\newcommand{\bea}{\begin{eqnarray}}
\newcommand{\eea}{\end{eqnarray}}
\newcommand{\bean}{\begin{eqnarray*}} 
	\newcommand{\eean}{\end{eqnarray*}}
\newcommand{\bel}{\begin{align}} 
\newcommand{\eel}{\end{align}}
\newcommand{\beln}{\begin{align*}} 
\newcommand{\eeln}{\end{align*}}
\newcommand{\bit}{\begin{itemize}}
	\newcommand{\eit}{\end{itemize}}

\makeatletter \@addtoreset{equation}{section}

\@addtoreset{Definition}{section}

\@addtoreset{Theorem}{section}

\@addtoreset{Proposition}{section}

\@addtoreset{Property}{section}

\@addtoreset{Assumption}{section}

\@addtoreset{Corollary}{section}

\@addtoreset{Lemma}{section}

\@addtoreset{Remark}{section}

\@addtoreset{Example}{section}

\@addtoreset{Condition}{section}

\newcommand{\No}[1]{\left|#1\right|}     

\def \E{\mathbb{E}}

\def \H{\mathbb{H}}
\def \L{\mathbb{L}}
\def \M{\mathbb{M}}
\def \N{\mathbb{N}}
\def \P{\mathbb{P}}
\def \Q{\mathbb{Q}}
\def \R{\mathbb{R}}

\def \Z{\mathbb{Z}}

\def \G{\mathbb{G}}

\def\={\;=\;}
\def\.{\;.}

\def\1{{\bf 1}}

\def\normeL2#1{\left\|{#1}\right\|_{L^2}}

\newcommand{\alias}[2]{
	\providecommand{#1}{}
	\renewcommand{#1}{#2}
}

\alias{\P}{\mathbb{P}}
\alias{\N}{\mathcal{N}}
\alias{\L}{\mathcal{L}}
\alias{\Z}{\mathbb{Z}}
\alias{\Q}{\mathbb{Q}}
\alias{\R}{\mathbb{R}}
\alias{\C}{\mathcal{C}}
\alias{\T}{\mathbb{T}}
\alias{\E}{\mathbb{E}}
\alias{\H}{\mathcal{H}}
\alias{\1}{\mathbf{1}}
\alias{\B}{\mathcal{B}}
\alias{\M}{\mathcal{M}}
\alias{\G}{\mathcal{G}}
\alias{\Y}{Y_{\bullet}}

\newcommand{\nc}{\newcommand}
\nc{\cA}{{\mathcal A}} \nc{\cB}{{\mathcal B}} \nc{\cC}{{\mathcal
		C}} \nc{\cD}{{\mathcal D}} \nc{\bbD}{\mathbb{D}}
\nc{\cG}{{\mathcal G}} \nc{\cF}{{\mathcal F}} \nc{\cS}{{\mathcal
		S}} \nc{\cU}{{\mathcal U}} \nc{\cH}{{\mathcal H}}
\nc{\cK}{{\mathcal K}} \nc{\cM}{{\mathcal M}} \nc{\cO}{{\mathcal
		O}} \nc{\cP}{{\mathcal P}} \nc{\bbE}{\mathbb{E}}
\nc{\bbEP}{\mathbb{E}_{\mathbb{P}}}\nc{\bbL}{\mathbb{L}}
\nc{\bbP}{\mathbb{P}} \nc{\bbQ}{\mathbb{Q}} \nc{\del}{\partial}
\nc{\Om}{\Omega} \nc{\om}{\omega} \nc{\bbR}{\mathbb{R}}
\nc{\bbC}{\mathbb{C}} 
\nc{\dXt}{\delta q_{t}}
\nc{\dXs}{\delta q_{s}} \nc{\bs}{\blacksquare} \nc{\dX}{\delta q}
\nc{\dY}{\Delta Y}
\nc{\dnkx}{\left(X(T^{n}_{k})-X(T^{n}_{k-1})\right)}
\nc{\esssup}{\mathrm{ess}\mbox{ }\mathrm{sup}}
\nc{\essinf}{\mathrm{ess}\mbox{ } \mathrm{inf}}
\nc{\dhats}{\widehat{\delta_s}}

\nc{\chf}{\mbox{$\mathbf1$}}
\nc{\ind}{\mathds{1}}

\begingroup\expandafter\expandafter\expandafter\endgroup
\expandafter\ifx\csname pdfsuppresswarningpagegroup\endcsname\relax
\else
\fi
\graphicspath{{./fig/}}

\nc{\mum}{ \mu_{\rm m} }
\nc{\muv}{ \mu_{\rm v} }
\nc{\mumv}{ \mu_{\rm mv} }
\nc{\Hm}{ H_{\rm m} }
\nc{\Hv}{ H_{\rm v} }

\usepackage{import}

\usepackage[onehalfspacing]{setspace}
\newcommand{\uproman}[1]{\uppercase\expandafter{\romannumeral#1}}

\usepackage{bbm}
\newcommand{\abstand}{\hspace{0pt}}
\newcommand\independent{\protect\mathpalette{\protect\independenT}{\perp}}
\def\independenT#1#2{\mathrel{\rlap{$#1#2$}\mkern2mu{#1#2}}}
\usepackage{graphicx}
\usepackage{float}

\usepackage{tabularx}
\usepackage{booktabs}
\usepackage{footnote}

\usepackage{caption}
\usepackage{subcaption}
\usepackage[skip=10pt]{subcaption}
\usepackage{ulem}

\usepackage{todonotes}
\newcounter{todocounter}

\newcommand\norm[1]{\left\lVert#1\right\rVert}
\usepackage{tikz}
\usepackage[bottom]{footmisc}
\usepackage[absolute]{textpos}
\usepackage{arydshln}
\newcommand\Eqref[1]{Equation~\eqref{#1}}
\crefname{subsection}{section}{sections}
\Crefname{subsection}{Section}{Sections}
\crefname{Example}{example}{examples}
\Crefname{Example}{Example}{Examples}
\Crefname{Assumption}{Assumption}{Assumptions}
\Crefname{AssumptionAppendix}{Assumption}{Assumptions}
\Crefname{Type}{Type}{Types}

\usepackage{breakcites}
\usepackage{bm}

\usepackage[pageref]{backref}
\renewcommand*{\backref}[1]{}
\renewcommand*{\backrefalt}[4]{%
	\ifcase #1 (Not cited.)%
	\or        (Cited on page~#2.)%
	\else      (Cited on pages~#2.)%
	\fi}

\begin{document}
\sloppy
	
\title{The Market Price of Jump Risk for Delivery Periods:\\Pricing of Electricity Swaps with Geometric Averaging\thanks{We would like to thank Christa Cuchiero for her fruitful comments and suggestions. Financial support from the Deutsche Forschungsgemeinschaft (DFG, German Research Foundation) – SFB 1283/2 2021 – 317210226 is gratefully acknowledged.}}

\author{
	Annika Kemper\thanks{Center for Mathematical Economics (IMW) at Bielefeld University,
	\href{mailto:annika.kemper@uni-bielefeld.de}{annika.kemper@uni-bielefeld.de}.} 
	\quad 
	Maren Diane Schmeck\thanks{Center for Mathematical Economics (IMW) at Bielefeld University, 
	\href{mailto:
			maren.schmeck@uni-bielefeld.de}{
			maren.schmeck@uni-bielefeld.de}.}
		}
\maketitle
\begin{abstract}
	\textit{In this paper, we extend the \textit{market price of risk for delivery periods} (MPDP) of electricity swap contracts by introducing a dimension for jump risk.
	As introduced by \cite{Kemper2022}, the MPDP arises through the use of geometric averaging while pricing electricity swaps in a geometric framework.	
	We adjust the work by \cite{Kemper2022} in two directions:
	First, we examine a \citeauthor{Merton1976} type model taking jumps into account.
	Second, we transfer the model to the physical measure by implementing mean-reverting behavior.
	We compare swap prices resulting from the classical arithmetic (approximated) average to the geometric weighted average.
	Under the physical measure, we discover a decomposition of the swap's market price of risk into the classical one and the MPDP.
}
	

\end{abstract}
	
	
	

\section{Introduction} \label{sec:introduction}
With the turn of the millennium, pricing derivatives on electricity has become important through the liberalization of energy markets.
Nowadays, new challenges appear due to the transition to a climate neutral energy system:
Electricity generated from renewable energy sources, like wind and solar energy, clearly depends on the  weather conditions of the season. Consequently, a rising share of renewable energy induces stronger intermittency and seasonality effects influencing especially delivery-dependent pricing effects.
In electricity markets, such delivery-dependent futures contracts are the most important derivatives. They deliver the underlying over a period of time since electricity is not storable on a large scale. We therefore call them electricity \textit{swaps}.
The dependence on the delivery time affects the price dynamics, the pricing measure, and the swap's market price of risk for delivery periods (MPDP)  introduced by \cite{Kemper2022}.
In this paper, we provide an extension of the MPDP. 
To do so, we adjust the model to a Merton type model taking jumps into account.
In addition, under the physical measure, we  identify a decomposition of the market price of risk into the classical one and the MPDP.\\


The delivery period is a unique feature of electricity markets that differs from other commodities such as oil, gas, or corn.
In fact, it plays a crucial role in the pricing of electricity swaps.
Following the market model approach, the electricity swap price results from averaging an instantaneous stream of futures with respect to the delivery time.
This approach goes back to the famous model by \cite{HJM}.
It was firstly connected to energy-related derivatives by \cite{ClewlowStrickland1999} and to electricity derivatives by \cite{Bjerksund} followed by a row of works (see, e.g., \Cref{tab:ClassificationSwapPriceModels} for geometric settings, \cite{Hinderks2020} for a structural model and \cite{CuchieroPersioGuidaSvalute2022} for measure-valued processes).
One stream of literature investigates spot based price dynamics and derive electricity futures based on the spot price referring to the day ahead market (see, e.g., \cite{CarteaFiguora2005, Cartea2008, Escribano2011}). In this paper, we focus on a HJM-type approach modelling the futures market directly. That is we consider so-called \textit{atomic} swap contracts inducing a delivery period of a month, that are used to price overlapping swap contracts delivering for example over a quarter or a year. We refer to \cite{Benth2008,BenthKoekebakker2008} and \cite{Kemper2022} for a construction of overlapping swap contracts based on atomic swaps. \\

The delivery period can be incorporated in different ways of averaging.
We distinguish between three types of averaging: Arithmetic, approximated, and geometric averaging.
\textit{Arithmetic averaging} is the classical way to implement the swap's delivery period and is convenient for arithmetic price dynamics. In particular, continuous arithmetic averaging is applied by  \cite{Benth2007}, \cite{Benth2008}, \cite{Benth2014}, \cite{Latini2019}, \cite{Benth2019}, and \cite{KleisingerYu}, among others (see also \Cref{tab:ClassificationSwapPriceModels}). For discrete arithmetic averaging, we refer to \cite{LuciaSchwartz2002} and \cite{BurgerMuller2004}.
Instead, arithmetic averaging of geometric price dynamics  is poorly suited since the resulting swap price dynamics are neither geometric nor Markovian.
It requires, e.g., an approximation of the swap price volatility introduced by \cite{Bjerksund} whenever we want to consider tractable swap price dynamics (see also \cite{Benth2008}, \cite{BenthKoekebakker2008}).
We call this procedure \textit{approximated averaging}.
\textit{Geometric averaging}, instead, does not require  any  approximations whenever the price dynamics are of geometric type and lead to suitable geometric dynamics (see \cite{Kemper2022}).
Hence, the geometric average is tailor-made for relative growth rate models. Nevertheless, the geometric average does not preserve the martingale property. This issue is tackled by \cite{Kemper2022} using a measure change with their MPDP.
Usually, negative prices are not observable in the data of the futures prices, such that
we stick to a geometric setting and compare the latter averaging procedures while adjusting the MPDP to a \citeauthor{Merton1976} type model.\\

\setlength{\extrarowheight}{0.18cm}
\begin{table}[htb] 
	\centering
	\begin{tabular}{|l|l|}
		\hline 
		& \textbf{Geometric Price Dynamics} \\[1ex]
		\hline
		\textbf{Arithmetic Average} 
		& \cite{KoekebakkerOllmar}\\ 
		& \cite{BenthKoekebakker2008}\\[1ex]
		\textbf{Approximated Average}  
		&\cite{Bjerksund}\\
		&\cite{Benth2008}\\[1ex]
		\textbf{Geometric Average} 
		& \cite{Kemper2022}\\
		\hline
	\end{tabular}
	\caption{Classification of selected electricity swap price models.}
	\label{tab:ClassificationSwapPriceModels}
\end{table}%

\begin{figure}[H]
	\centering
	\begin{tikzpicture}
	\path (0,0) coordinate[label=left:$\mathbb{P}$]  (A) --(20:4) coordinate[label=above:$\widetilde{\mathbb{Q}}$] (C) -- ++(-20:4) coordinate[label=right:$\mathbb{Q}$] (B);
	\draw[-latex] (A) -- (B);
	\draw[-latex] (A) -- (C);
	\draw[-latex] (B) -- (C);
	\path (A) -- (B) node[midway,below] {$\Pi^{\mathbb{P}\mathbb{Q}}$};
	\path (A) -- (C) node[midway,above] {$\Pi^{\mathbb{P}\widetilde{\mathbb{Q}}}$};
	\path (C) -- (B) node[midway,above] {$\Pi^{\mathbb{Q}\widetilde{\mathbb{Q}}}$};
	\end{tikzpicture}
	\caption{Considered measure changes between the physical measure $\mathbb{P}$, the artificial risk-neutral measure $\mathbb{Q}$, and the swap's pricing measure $\widetilde{\mathbb{Q}}$ as well as their connections with the true market prices of risk $\Pi^{\mathbb{P}\mathbb{Q}}$, the classical market price of risk $\Pi^{\mathbb{P}\widetilde{\mathbb{Q}}}$, and the MPDP denoted by$\Pi^{\mathbb{Q}\widetilde{\mathbb{Q}}}$.}
	\label{fig:OverviewMeasureChanges}
\end{figure}
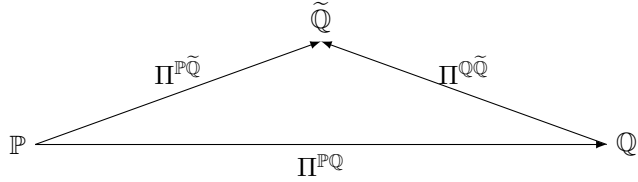

Both papers, \cite{Kemper2022} and \cite{Bjerksund},
investigate the modeling of the delivery period explicitly through a continuous weighted averaging approach for geometric futures prices. 
Both approaches lead to Markovian and geometric swap price dynamics.
We discuss similarities and differences between these approaches and introduce a numéraire caused by the different averaging techniques in \Cref{sec:AVERAGING}. 
In line with the market model approach, we base the averaging procedure on a continuous stream of futures contracts that is a martingale under the futures risk-neutral measure $\mathbb{Q}$. 
We consider electricity futures with instantaneous delivery as artificial contracts and we therefore refer to $\mathbb{Q}$ as the \textit{artificial measure}. 
The resulting swap price dynamics based on geometric averaging are not a martingale under $\mathbb{Q}$. We then define the MPDP of diffusion and jump risk and a new pricing measure~$\widetilde{\mathbb{Q}}$, which can thus be used to price derivatives on the swap.
We may refer  to $\widetilde{\mathbb{Q}}$ as the ``correct'' or ``true'' risk-neutral measure since the swap price is a $\widetilde{\mathbb{Q}}$-martingale without any approximations.
Under the artificial measure, the swap based on the approximated version is directly a martingale.
Therefore, we  call $\mathbb{Q}$ also the ``classical'' risk-neutral measure.
It is a clear advantage that the approximated average preserves the  martingale property of the swap under the measure~$\mathbb{Q}$.
A decomposition of the market price of risk for electricity swaps arises when turning to the physical measure $\mathbb{P}$. \Cref{fig:OverviewMeasureChanges} gives an overview over the connections between the different measures $\mathbb{P}$, $\mathbb{Q}$, and $\widetilde{\mathbb{Q}}$ and the true market price of risk $\Pi^{\mathbb{P}\mathbb{Q}}$, the classical market price of risk $\Pi^{\mathbb{P}\widetilde{\mathbb{Q}}}$, and the MPDP denoted by $\Pi^{\mathbb{Q}\widetilde{\mathbb{Q}}}$.\\

Indeed, the MPDP is triggered by typical features of the electricity market entering the swap's volatility.
In particular, delivery-dependent effects like seasonalities and term-structure effects play a crucial role.
\cite{fanelli2019seasonality} empirically identify seasonalities in the swap's delivery period by considering implied volatilities of electricity options.
Renewable energy, like wind and solar energy, intensify especially the seasonal effects mentioned before. 
An additional property of electricity and commodity markets is the Samuelson effect (see \cite{Samuelson}): The closer we reach the end of the maturity, the more effect the volatility has.
\cite{BenthParaschiv} and \cite{JaeckLautier} provide empirical evidence for the Samuelson effect in the volatility term-structure of electricity swaps. It can also be  observed in the implied volatilities of electricity options, especially far out and in the money (see \cite{kiesel}).
\cite{Kemper2022} characterize the MPDP for such seasonalities and term-structure effects within a stochastic volatility model through the variance per unit of expectation of the delivery-dependent effects.
We contribute to the literature by investigating the MPDP analytically, affected by seasonalities and the Samuelson effect. Moreover, we lay the foundation for the empirical analysis of the MPDP.\\

Further characteristics of the electricity swap market are mean-reversion and jump behavior.
As mentioned by \cite{Latini2019} and \cite{KleisingerYu} among others, mean-reversion is an important property of the electricity swap prices. 
\cite{KoekebakkerOllmar} empirically validate that the short-term price varies around the long-term price, which confirms mean-reverting behavior.
As \cite{Benth2019}, we face the problem of changing a mean-reverting process to the risk-neutral measure.
We extend their measure change to the geometric setting. We even provide a proof  for stochastic volatility settings in order to address models such as \cite{Kemper2022} and \cite{schneider2018samuelson}.
Besides mean-reversion, \cite{Benth2019} include jumps as an outstanding characteristic of electricity prices. 	In particular, they consider compound Poisson processes under the physical measure in a mean-reverting, arithmetic setting.
While adjusting the paper by \cite{Kemper2022} to jumps, we establish the \textit{MPDP of jump risk} whenever the jump coefficient relies on delivery-dependent effects. \\

The contribution to the literature is twofold: 
First, we adjust the paper by \cite{Kemper2022} to the jump case under the artificial risk-neutral measure leading to an extended characterization of the MPDP regarding diffusion \textit{and} jump risk. Hence, even if approximated averaging in the spirit of \cite{Bjerksund} is performed in a Merton type model, the model can easily overcome any approximation issues through an application of the MPDP.
Second, we transfer the model to the physical measure and compare the swap prices resulting from geometric  and approximated  averaging
as well as their risk-neutral measures revealing the decomposition of the market price of risk into the classical one and the MPDP. Consequently, the model lays the foundation for empirical investigations in the future. \\

The paper is organized as follows: 
\Cref{sec:AVERAGING} presents the geometric averaging approach under the artificial risk-neutral measure applied to the jump-type futures curve. In addition, it presents the  MPDP of diffusion and jump risk. 
\Cref{sec:Model} introduces the model under the physical measure and identifies the decomposition of the market price of risk.
Finally, \Cref{sec:summary} concludes our main findings.

\section{On the MPDP of Diffusion and Jump Risk} \label{sec:AVERAGING} 
We particularly focus on an electricity swap contract delivering  1 MWh of electricity during the agreed delivery period $(\tau_1,\tau_2]$. 
At a trading day $t\leq \tau_1$ before the contract expires,  we denote the swap price by $F(t, \tau_1,\tau_2)$ settled such that the contract is entered at no cost. 
It can be interpreted as an average price of instantaneous delivery. Motivated by this interpretation, we consider an artificial futures contract with price $f(t,\tau)$ that stands for instantaneous delivery at time $\tau\in(\tau_1,\tau_2]$. Note that such a contract does not exist on the market  but it turns out to be useful for modeling purposes when considering delivery periods (see, e.g., \cite{Benth2019} and \cite{Kemper2022}). \\

Following the approach by \cite{HJM}, we derive the price of an electricity swap contract based on an instantaneous futures price model.
More precisely, we compare two types of swap prices resulting from geometric and approximated averaging. The goal of this section is to investigate the pricing spread between both approaches in order to quantify the consequences of the approximation and thus the effect of the precise geometric averaging procedure. As the pricing spread goes along with different risk-neutral measures, we additionally investigate the distance of both risk-neutral measures quantified by the MPDP. Moreover, we characterize the MPDP for specific volatility functions and different jump size distributions.


\paragraph{The Model.}
Consider a filtered probability space $(\Omega, \mathcal{F}, (\mathcal{F}_t)_{t\in[0,\tau]},\mathbb{Q})$,
where the filtration satisfies the usual conditions.
We first model the solution of a futures contract and then derive the corresponding dynamics to avoid lacks of existence in the presence of jumps (see \cite{Papapantoleon}). At time $t\leq\tau$, let the logarithmic price process of the futures contract be defined as
	\begin{align}
	\ln f(t,\tau) &= \ln f(0,\tau)+ \int_{0}^{t}\sigma(s,\tau)dW^{\mathbb{Q}}_s
	+\int_{0}^{t}\eta(s,\tau)  d\widetilde{J}^{\mathbb{Q}}_s - \int_{0}^{t}c^{\mathbb{Q}}(s,\tau)ds\;, \label{eq:futuresQsol1}
	\end{align}
 with initial non-random conditions $f(0,\tau) >0$.	
	Moreover, $W^{\mathbb{Q}}$ is a one-dimensional standard Brownian motion under $\mathbb{Q}$ independent of the jump process $\widetilde{J}^{\mathbb{Q}}$. 
	In particular, $\widetilde{J}^{\mathbb{Q}}$ is a compound compensated jump process defined through the compensated Poisson random measure $\widetilde{N}^{\mathbb{Q}}(dt,dz)=N(dt,dz)-\ell^{\mathbb{Q}}(dz)dt$:
	\begin{align}
	\widetilde{J}^{\mathbb{Q}}_t=\int_{0}^{t}\int_{\mathbb{R}}z\widetilde{N}^{\mathbb{Q}}(ds,dz)\;,
	\end{align}
	with Lévy measure $\ell^{\mathbb{Q}}(dz)=\lambda^{\mathbb{Q}} G(dz)$, which is independent of the delivery time, and where $\lambda^{\mathbb{Q}}>0$ indicates the jump intensity and $G(dz)$ the jump size distribution.
	The last term in \Eqref{eq:futuresQsol1} defines the compensator of the logarithmic return under the current measure $\mathbb{Q}$:
	\begin{align}
	c^{\mathbb{Q}}(t,\tau)=\frac12\sigma^2(t,\tau)+ \psi^{\mathbb{Q}}(i\eta(t,\tau))\;, \label{eq:compensator}
	\end{align}
	where $\psi^{\mathbb{Q}}(ir)$ is the integrand of the Lévy-Khintchine exponential defined through the moment generating function 
	\begin{align}
	\psi^{\mathbb{Q}}(r):=\int_{\mathbb{R}}\left(e^{rz}-1-rz\right)\ell^{\mathbb{Q}}(dz)\;.
	\end{align}
	We assume that the futures price volatility and jump coefficients, $\sigma(t,\tau)$ and $\eta(t,\tau)$, are deterministic and that the futures price $f(t,\tau)$  is $\mathcal{F}_t$-adapted for $t\in[0,\tau]$. We further assume that they satisfy suitable integrability and measurability conditions (see  \Cref{ass:TechnicalRequirementsQ1} in Appendix \ref{app:tech_requ3} for details) to ensure that the process in \Eqref{eq:futuresQsol1} is a $\mathbb{Q}$-martingale, and that \Eqref{eq:futuresQsol1} gives the unique solution to the process evolving as
	\begin{align}
	\frac{df(t,\tau)}{f(t-,\tau)}= \sigma(t,\tau)dW^{\mathbb{Q}}_t+\int_{\mathbb{R}}\left(e^{\eta(t,\tau)z}-1\right)\widetilde{N}^{\mathbb{Q}}(dt, dz)\;. \label{eq:futuresQ}
	\end{align}
As $\sigma(t,\tau)$ depends on both, trading time $t$ and delivery time $\tau$, we allow for volatility structures as the Samuelson effect or seasonalities in the delivery time, which are addressed in \Cref{Ex:Seasonality,Ex:Samuelson}.

\paragraph{Implementing the Delivery Period.}
Following the Heath-Jarrow-Morton approach to price futures and swaps in electricity markets, the swap price is usually defined as the \textit{arithmetic weighted average} of futures prices (see, e.g., \cite{Benth2008},  \cite{Bjerksund}, and \cite{Benth2019}):
\begin{equation} \label{eq:arithmeticnoarbitrage}
F^A(t,\tau_1,\tau_2):=\int_{\tau_1}^{\tau_2}w(u, \tau_1, \tau_{2})f(t,u)du\;, 
\end{equation} 
for a general weight function
\begin{align}
w(u,\tau_1, \tau_2):=\frac{\hat{w}(u)}{\int_{\tau_1}^{\tau_2}\hat{w}(v)dv}\;, ~~~~ \text{for } u\in(\tau_1,\tau_2]\;, \label{eq:weights}
\end{align}
where $\hat{w}(u)>0$ is the corresponding settlement function.
Note that $w$ defines a probability density function with support on $(\tau_1,\tau_2]$ since it is positive and integrates to one, that is $\int_{\tau_1}^{\tau_2}w(u,\tau_1, \tau_2)du=1$.
Hence, we denote $U$ as a random delivery variable with density $w(u,\tau_1,\tau_2)$ (see also \cite{Kemper2022}).
The most popular example is given by a constant settlement type $\hat{w}(u) =1$, such that the density becomes $w(u,\tau_1, \tau_2)= \frac{1}{\tau_2 - \tau_1}$ and $U\sim \mathcal{U}((\tau_1,\tau_2])$ is uniformly distributed over the delivery period. This corresponds to a one-time settlement. A~continuous settlement over the time interval $(\tau_1,\tau_2]$ is covered by a continuous discount function $	\hat{w}(u) =e^{-ru}$, where $r$ is the constant interest rate (see, e.g., \cite{Benth2008}). \\

The arithmetic average of the futures price as in \Eqref{eq:arithmeticnoarbitrage} leads to tractable dynamics for the swap as long as one  assumes an arithmetic structure of the futures prices as well. 
This is based on the fact that arithmetic averaging is tailor-made for absolute growth rate models.
Nevertheless,  if one defines the futures price as a geometric process as in \Eqref{eq:futuresQ}, one can show that the dynamics of the swap price $F^A$ defined through \Eqref{eq:arithmeticnoarbitrage} are given by
\begin{equation}
\begin{split}
\frac{dF^A(t,\tau_1,\tau_2)}{F^A(t-,\tau_1,\tau_2)}=&~\Big[\sigma(t,\tau_2) -\int_{\tau_1}^{\tau_2}\frac{\partial \sigma}{\partial u}(t,u)\frac{w(\tau, \tau_1,\tau_2)}{w(\tau, \tau_1, u)} \frac{F^A(t, \tau_1,u)}{F^A(t, \tau_1,\tau_2)} du\Big] ~dW^\mathbb{Q}_t\\
&~+ \int_{\mathbb{R}}\Big(e^{\eta(t,\tau_2)z}-1- \int_{\tau_1}^{\tau_2}\frac{\partial e^{\eta(s,u)z}}{\partial u}\frac{w(\tau, \tau_1,\tau_2)}{w(\tau, \tau_1, u)}\frac{F^A(t, \tau_1,u)}{F^A(t, \tau_1,\tau_2)} du\Big)~\widetilde{N}^{\mathbb{Q}}(dz,dt)\;,
\end{split}
\end{equation}
for any $\tau\in(\tau_1,\tau_2]$ (see \cite{Benth2008}, cf.\ Chapter 6.3.1). Thus, the dynamics of the swap price is neither a geometric process nor Markovian, which makes it unhandy for further analysis.  
To overcome this issue, \cite{Bjerksund} suggest an approximation in the setup without jumps, which we call \textit{approximated averaging} since it is the arithmetic average of approximated logarithmic returns. Approximated averaging maintains the martingale property
meaning that the swap is a martingale whenever $f$ is a martingale. 
If we transfer the approximated averaging procedure to our jump setting, we can define the  swap price process based on approximated averaging by
\begin{align}
\frac{dF^a(t,\tau_1,\tau_2)}{F^a(t-,\tau_1,\tau_2)}:=\int_{\tau_1}^{\tau_2}w(u,\tau_1,\tau_2) \frac{df(t,u)}{f(t-,u)} du\;. \label{eq:arithAverage_drift}
\end{align}
In contrast, \textit{geometric averaging} originates from the arithmetic average of logarithmic returns without any need for approximations.
Hence, in line with \cite{Kemper2022}, we define the swap price originating from geometric averaging by
\begin{align}
F(t,\tau_1,\tau_2):=e^{\int_{\tau_1}^{\tau_2}w(u,\tau_1,\tau_2)\ln f(t,u)du}\;, \label{eq:GeomAverage}
\end{align}
(see also \cite{KemnaVorst1990}).
Assume that the volatility and jump coefficients satisfy further integrability conditions (see \Cref{ass:TechnicalRequirementsQ2} in \Cref{app:tech_requ3}).
It turns out, that the resulting swap price dynamics is a geometric process with a non-zero drift term:
\begin{Lemma}[The Swap Price under $\mathbb{Q}$] \label{lemma:dyn_Swap_Q}
	Let  \Cref{ass:TechnicalRequirementsQ2} in \Cref{app:tech_requ3} be satisfied.
	Under the artificial pricing measure $\mathbb{Q}$, the dynamics of the swap price process $F(\cdot,\tau_1, \tau_2)$, defined in \Eqref{eq:GeomAverage}, are given by
	\begin{equation}\label{eq:FunderQ}
	\begin{split}
	\frac{d F(t,\tau_1,\tau_2)}{F(t-,\tau_1,\tau_2)} =&~
	\mathbb{E}[\sigma(t,U)]~dW^{\mathbb{Q}}_t+\int_{\mathbb{R}} \left(e^{\mathbb{E} [\eta(t,U)]z}-1\right)\widetilde{N}^{\mathbb{Q}}(dt,dz)\\
	&~-\left(\frac{1}{2}\mathbb{V} \left[\sigma(t,U) \right]
	+\mathbb{E} [\psi^{\mathbb{Q}}(\eta(t,U))]-\psi^{\mathbb{Q}}(\mathbb{E} [\eta(t,U)]) \right) dt\;,
	\end{split}
	\end{equation}
	where $U$ denotes the random delivery variable  with density $w(u,\tau_1,\tau_2)$.
\end{Lemma}
\begin{proof}
	Plugging the integral representation of the futures rate process from \Eqref{eq:futuresQsol1} into \Eqref{eq:GeomAverage} 
  gives us $F(t,\tau_1,\tau_2)
	=F(0, \tau_1, \tau_{2}) e^{\bar{X}(t,\tau_1,\tau_2)}$,
 where an application of the stochastic Fubini Theorem (see \cite{Protter2005}, cf.\ Theorem~65, Chapter IV.6) leads to
	\begin{align}
	\bar{X}(t,\tau_1,\tau_2)=&\int_{0}^{t}\mathbb{E} \Big[\sigma(s,U)\Big]dW^{\mathbb{Q}}_s+\int_{0}^{t}\mathbb{E} [\eta(s,U)]d\widetilde{J}^{\mathbb{Q}}_s-\frac12\int_{0}^{t}\mathbb{E} \Big[\sigma^2(s,U)\Big]ds
	-\int_{0}^{t}
	\mathbb{E} [\psi^{\mathbb{Q}}( \eta(s,U))]ds\;. \label{eq:Xbar}
	\end{align}
	Then, \Eqref{eq:FunderQ} follows using Itô's formula (see, e.g., \cite{OksendalSulem}). \qed\\
\end{proof}

Having presented the three procedures of continuous time averaging that are used to derive the swap from an underlying artificial futures curve, we would like to compare them:
Arithmetic averaging, defined by \Eqref{eq:arithmeticnoarbitrage}, is tractable for arithmetic futures curves, whereas approximated averaging, defined by \Eqref{eq:arithAverage_drift}, and geometric averaging, defined by \Eqref{eq:GeomAverage}, are well suited for geometric futures curves.
In line with a series of literature (see \Cref{tab:ClassificationSwapPriceModels}), we follow the geometric approach.
Our goal throughout this paper is to investigate the pricing spread between geometric and approximated averaging analytically.

\paragraph{The MPDP.}
Although the futures price $f$ and the approximated $F^a$ are martingales under the pricing measure $\mathbb{Q}$, the swap price $F$ is not a $\mathbb{Q}$-martingale: Indeed, the swap price process under $\mathbb{Q}$ has a negative drift term consisting of two parts given by the swap's variance and the difference between the averaged Lévy-Khintchine integrand and the Lévy-Khintchine integrand of the averaged jump coefficient.
Hence, using geometric averaging leads to a new interpretation of risk related to the delivery period as we will analyze in the following.\\

Analogous to \cite{Kemper2022}, we derive the corresponding risk-neutral measure $\widetilde{\mathbb{Q}}$ under which the electricity swap price $F$ is a martingale.
For deriving the swap's risk-neutral measure, we thus define the MPDP extended to jumps in the following.
\begin{Definition}[The MPDP] \label{Def_MPDP}
At time $t\in[0,\tau_1]$, the \textit{market price of diffusion and jump risk for delivery periods} associated to the delivery period $(\tau_1,\tau_2]$ is defined by
$\Pi^{\mathbb{Q}\widetilde{\mathbb{Q}}}:=(\Pi_1^{\mathbb{Q}\widetilde{\mathbb{Q}}},\Pi_2^{\mathbb{Q}\widetilde{\mathbb{Q}}})$, where
\begin{align}
&\Pi_1^{\mathbb{Q}\widetilde{\mathbb{Q}}}(t,\tau_1,\tau_2) :=-\frac12 \frac{\mathbb{V} \left[\sigma(t,U)\right]}{\mathbb{E} \left[\sigma(t,U)\right]}\;, \label{eq:AddMP1}\\
&\Pi_2^{\mathbb{Q}\widetilde{\mathbb{Q}}}(t, \tau_1,\tau_2):= 
- \frac{\int_{\mathbb{R}}\mathbb{E} [e^{\eta(t,U)z}]-e^{\mathbb{E} [\eta(t,U)]z}\ell^{\mathbb{Q}}(dz)}{\int_{\mathbb{R}}\left(e^{\mathbb{E} [\eta(t,U)]z}-1\right)\ell^{\mathbb{Q}}(dz)}\;.\label{eq:AddMP2}
\end{align}
\end{Definition}
In general, the MPDP does not coincide with the market price of risk. In fact, it is an additional risk that has to be taken into account whenever approximated averaging is conducted.
Technically speaking, the MPDP characterizes the distance between the martingale measure of the swaps, $F$ and $F^a$, resulting from geometric and approximated averaging.
In particular, $\Pi_1$ refers to the additional diffusion risk, which is measurable and $\mathcal{F}_t$-adapted as $\sigma(t,u)$ is. 
It can be interpreted as the trade-off between the weighted average variance of a stream of futures, on the one hand, and the variance of the swap, on the other hand (see also
\cite{Kemper2022} for an elaboration of the MPDP $\Pi_1^{\mathbb{Q}\widetilde{\mathbb{Q}}}$ and a detailed interpretation).
$\Pi_2^{\mathbb{Q}\widetilde{\mathbb{Q}}}$ is the additional jump risk, which is the difference between the Lévy-Khintchine integrands standardized by the swap's jump coefficient.
\begin{Remark} \label{re:1}
	\begin{itemize}
	\item[$(i)$] Note that $\Pi^{\mathbb{Q}\widetilde{\mathbb{Q}}}$ would be zero, whenever the volatility and jump coefficients are independent of delivery time. For this reason, we call  $\Pi^{\mathbb{Q}\widetilde{\mathbb{Q}}}$ the \textit{market price of  risk for delivery periods} (MPDP).
	\item[$(ii)$] The MPDP of diffusion and jump risk is strengthened by delivery-dependent effects within the volatility and jump coefficients. For example, pronounced term-structure effects or seasonalities in the delivery period within these coefficients capture a distinct dependence on the delivery period and, consequently, lead to a high MPDP (see also \Cref{Ex:Seasonality,Ex:Samuelson} for the MPDP of diffusion risk and \Cref{ex:JumpsNormal,ex:JumpsExponential}.
	\item[$(iii)$] $\Pi_1^{\mathbb{Q}\widetilde{\mathbb{Q}}}$ is in line with the MPDP for diffusion risk found in \cite{Kemper2022}, where a stochastic volatility scenario is considered.
	\end{itemize}
\end{Remark} 

In a next step, we would like to characterize the MPDP of diffusion risk $\Pi^{\mathbb{Q}\widetilde{\mathbb{Q}}}_1$ more explicitly.
The MPDP of diffusion risk arises through delivery-dependent volatility effects such as seasonality in delivery periods and the Samuelson effect (see \cite{Kemper2022}). We state the corresponding MPDP in the following two examples while assuming a one-time settlement such that $w(t,\tau_1,\tau_2)=\frac{1}{\tau_2-\tau_1}$.

\begin{Example}[Seasonal Volatility]\label{Ex:Seasonality}
Inspired by \cite{fanelli2019seasonality}, we capture seasonality in the delivery period by incorporating a trigonometric function into the futures volatility  $\sigma(t,u)=S_1(u)$ in (see \Eqref{eq:AddMP1}) given by
\begin{align}
S_1(u):= a+ b \cos (2\pi (u+c))\;,
\end{align}
for $a>b>0$ and $c\in[0,1)$.
According to \Cref{Def_MPDP}, this leads to a MPDP of diffusion risk of the following form
\begin{align}
\Pi_1^{\mathbb{Q}\widetilde{\mathbb{Q}}}(t,\tau_1,\tau_2)=-\frac12\frac{\mathbb{V}[S_1(U)]}{\mathbb{E}[S_1(U)]}\;,
\end{align}
where 
\begin{align}
\mathbb{E}[S_1(U)]&= a+\frac{b}{2\pi (\tau_2-\tau_1)} \Big[\sin(2\pi(u+c))\Big]^{u=\tau_2}_{u=\tau_1}\;, \label{eq:Es}\\
\mathbb{E}[S_1(U)^2]&= a^2+\frac{b^2}{2}+\frac{ab}{\pi(\tau_2-\tau_1)}\Big[\sin(2\pi(u+c))\Big]^{u=\tau_2}_{u=\tau_1}+\frac{b^2}{8\pi(\tau_2-\tau_1)}\Big[\sin(4\pi(u+c))\Big]^{u=\tau_2}_{u=\tau_1}\;. \label{eq:Es2}
\end{align}
\end{Example}

\begin{Example}[Term-Structure Volatility]\label{Ex:Samuelson}
We implement the Samuelson effect as in \cite{schneider2018samuelson} into the futures volatility $\sigma(t,u)=S_2(u-t)$ (see \Eqref{eq:AddMP1}) through an exponential function with exponential damping factor $\Lambda>0$ and terminal volatility $\bar{\lambda}>0$ given by
\begin{align}
	S_2(u-t):=\bar{\lambda}e^{-\Lambda(u-t)}\;.
\end{align}
According to \Cref{Def_MPDP}, this leads to a MPDP of diffusion risk of the following form
\begin{align}
\Pi_1^{\mathbb{Q}\widetilde{\mathbb{Q}}}(t,\tau_1,\tau_2)=-\frac12\frac{\bar{\bar{\Lambda}}-\bar{\Lambda}^2}{\bar{\Lambda}}e^{-\Lambda(\tau_1-t)}\;,
\end{align}
for constant parameters $\bar{\Lambda}:=\frac{\bar{\lambda}(1-e^{-\Lambda(\tau_2-\tau_1)})}{\Lambda (\tau_2-\tau_1)}$ and $ \bar{\bar{\Lambda}}:=\frac{\bar{\lambda}^2(1-e^{-2\Lambda(\tau_2-\tau_1)})}{2\Lambda (\tau_2-\tau_1)}$ implicitly depending on the delivery period.
\end{Example}
Hence, the MPDP of diffusion risk is constant for a fixed contract in \Cref{Ex:Seasonality}, whereas the Samuelson effect remains still visible in \Cref{Ex:Samuelson}. For a detailed investigation of the volatility term structure, we refer to \cite{Kemper2022}. \\

The MPDP of jump risk is triggered by delivery-dependent jump effects.
For notational convenience, we choose $\eta(t,u)$ independent of trading time and again assume a one-time settlement.
In the following examples, we characterize the MPDP of jump risk and the spread based on two suitable jump size distributions: Normal and exponential.
We also state corresponding moments of the distributions following \cite{GrayPitts2012} (cf.\ Chapter~2).

\begin{Example}[Normal Jump Sizes] \label{ex:JumpsNormal}
 If the jump sizes are normally distributed with $Z\sim\mathcal{N}(\mu_J,\sigma_J^2)$, then the moment generating function is given by
 \begin{align}
 M_Z(\eta)=e^{\mu_J\eta+\frac12 \sigma_J^2\eta^2}\;,
 \end{align}
 such that the MPDP and the spread in Equations~\eqref{eq:AddMP2} and \eqref{eq:Spread2} are given by
	\begin{align}
	{\Pi}^{\mathbb{Q}\widetilde{\mathbb{Q}}}_2(\tau_1,\tau_2)  &= - \frac{\mathbb{E}[e^{\frac12 \eta^2(U)\sigma_J^2+\eta(U)\mu_J}]-e^{\frac12 \mathbb{E}[\eta(U)]^2\sigma_J^2+\mathbb{E}[\eta(U)]\mu_J}}{e^{\frac12 \mathbb{E}[\eta(U)]^2\sigma_J^2+\mathbb{E}[\eta(U)]\mu_J}-1}\;.
	\end{align}
	The fourth moment is attained by  $\int_{\mathbb{R}}z^4G(dz)=\mu_J^4+ 6\mu_J^2 \sigma_J^2+3\sigma_J^4$.
\end{Example}


\begin{Example}[Exponential Jump Sizes] \label{ex:JumpsExponential}
	If the jump sizes are exponentially distributed with
	$Z\sim\mathcal{E}xp(\lambda_J)$, for  $\lambda_J>0$, then the moment generating function is given by
	\begin{align}
	M_Z(\eta)=\left(1-\frac{\eta}{\lambda_J}\right)^{-1}\;,
	\end{align}
	for $\eta<\lambda_J$, such that the MPDP and the spread in Equations~\eqref{eq:AddMP2} and \eqref{eq:Spread2}  are given by
	\begin{align}
	{\Pi}^{\mathbb{Q}\widetilde{\mathbb{Q}}}_2(\tau_1,\tau_2)  &=- \frac{\lambda_J}{\mathbb{E}[\eta(U)]}\left(1-\mathbb{E}\left[\frac{\lambda_J-\mathbb{E}[\eta(U)]}{\lambda_J-\eta(U)}\right]\right) \;,
	\end{align}
	defined for $\eta(U)<\lambda_J$ and $\mathbb{E}[\eta(U)]<\lambda_J$.
	The $n$-th moment is attained by $\int_{\mathbb{R}}z^nG(dz)=\frac{n!}{\lambda_J^n}$ for $n\in\mathbb{N}$. 
\end{Example}

\paragraph{On the Swap's Martingale Measure.}
We define a new pricing measure $\widetilde{\mathbb{Q}}$, such that the swap price process $F(\cdot, \tau_1,\tau_2)$ is a martingale.
Following \cite{OksendalSulem}, define the Radon-Nikodym density through 
\begin{align}
Z^{\mathbb{Q}\widetilde{\mathbb{Q}}}(t,\tau_1,\tau_2) =\prod_{j=1}^{2}
Z_j^{\mathbb{Q}\widetilde{\mathbb{Q}}}(t,\tau_1,\tau_2) \;, \label{eq:densityPtilde}
\end{align}
where
\begin{align}
Z_1^{\mathbb{Q}\widetilde{\mathbb{Q}}}(t,\tau_1,\tau_2)&:=
e^{-\int_{0}^{t}\Pi_1^{\mathbb{Q}\widetilde{\mathbb{Q}}}(s,\tau_1,\tau_2)d\widetilde{W}^{\mathbb{Q}}(s)-\frac{1}{2}\int_{0}^{t}\Pi_1^{\mathbb{Q}\widetilde{\mathbb{Q}}}(s,\tau_1,\tau_2)^2ds}\;,\\
Z_2^{\mathbb{Q}\widetilde{\mathbb{Q}}}(t,\tau_1,\tau_2)&:=
e^{\int_{0}^{t}\int_{\mathbb{R}}\ln(1-\Pi_2^{\mathbb{Q}\widetilde{\mathbb{Q}}}(s, \tau_1,\tau_2))\widetilde{N}^{\mathbb{Q}}(ds,dz)+
	\int_{0}^{t}\int_{\mathbb{R}}\left(\ln(1-\Pi_2^{\mathbb{Q}\widetilde{\mathbb{Q}}}(s, \tau_1,\tau_2))+\Pi_2^{\mathbb{Q}\widetilde{\mathbb{Q}}}(s, \tau_1,\tau_2)\right) \ell^{\mathbb{Q}}(dz)ds}\;.
\end{align}
Assume that 
\begin{align}
\mathbb{E}_{\mathbb{Q}}[Z^{\mathbb{Q}\widetilde{\mathbb{Q}}}(\tau_1,\tau_1,\tau_2)]=1\;, \label{eq:MG_Property}
\end{align}
which means that $Z^{\mathbb{Q}\widetilde{\mathbb{Q}}}(\cdot,\tau_1,\tau_2)$ is indeed a martingale for the entire trading time.
We will show later that the martingale property is satisfied for suitable models such that \Eqref{eq:MG_Property} holds true.
We then define the new measure $\widetilde{\mathbb{Q}}$ through the Radon-Nikodym density
\begin{align}
\frac{d\widetilde{\mathbb{Q}}}{d\mathbb{Q}}=Z^{\mathbb{Q}\widetilde{\mathbb{Q}}}(\tau_1,\tau_1,\tau_2)\;,
\end{align}
which clearly depends on the delivery period $(\tau_1,\tau_2]$. Girsanov's theorem states that if we define the process $W^{\widetilde{\mathbb{Q}}}$ and the random measure $\widetilde{N}^{\widetilde{\mathbb{Q}}}(dt,dz)$ by
\begin{align}
dW^{\widetilde{\mathbb{Q}}}_t&= dW^{\mathbb{Q}}_t +\Pi_1^{\mathbb{Q}\widetilde{\mathbb{Q}}}(t,\tau_1,\tau_2)dt\;, \\
\widetilde{N}^{\widetilde{\mathbb{Q}}}(dt,dz)&= \widetilde{N}^{\mathbb{Q}}(dt,dz)+\Pi_2^{\mathbb{Q}\widetilde{\mathbb{Q}}}(t, \tau_1,\tau_2)\ell^{\mathbb{Q}}(dz)dt\;,
\end{align}
then $W^{\widetilde{\mathbb{Q}}}$ is a Brownian motion under $\widetilde{\mathbb{Q}}$ and $\widetilde{N}^{\widetilde{\mathbb{Q}}}(\cdot,\cdot)$ is the $\widetilde{\mathbb{Q}}$-compensated Poisson random measure of $N(\cdot,\cdot)$  with compensator $\left(1-\Pi_2^{\mathbb{Q}\widetilde{\mathbb{Q}}}(s, \tau_1,\tau_2)\right)\ell^{\mathbb{Q}}(dz)$. 
Under some further  assumptions, ensuring that $Z_2^{\mathbb{Q}\widetilde{\mathbb{Q}}}$ stays positive and that $Z^{\mathbb{Q}\widetilde{\mathbb{Q}}}$ is a true martingale (see \Cref{ass:TechnicalRequirementsQ3} in \Cref{app:tech_requ3}), a straightforward valuation leads  to the following result:
\begin{Proposition}[The Swap Price under $\widetilde{\mathbb{Q}}$] \label{prop:FunderPtilde}
	Let \Cref{ass:TechnicalRequirementsQ3} in \Cref{app:tech_requ3} be satisfied.
	The swap price process $F(\cdot,\tau_1,\tau_2)$, defined in~\eqref{eq:GeomAverage}, is a martingale under $\widetilde{\mathbb{Q}}$. The swap price dynamics are given by
	\begin{align}
	\frac{d F(t,\tau_1,\tau_2)}{F(t-,\tau_1,\tau_2)} = 
	\mathbb{E}[\sigma(t,U)]dW^{\widetilde{\mathbb{Q}}}_t+\int_{\mathbb{R}} \left(e^{\mathbb{E} [\eta(t,U)]z}-1\right)\widetilde{N}^{\widetilde{\mathbb{Q}}}(dt,dz)
	\;,
	\end{align}
	where $W^{\widetilde{\mathbb{Q}}}$ is a Brownian motion under $\widetilde{\mathbb{Q}}$ and $\widetilde{N}^{\widetilde{\mathbb{Q}}}(\cdot,\cdot)$ is the compound compensated Poisson random measure under $\widetilde{Q}$ with Lévy measure $\left(1-\Pi_2^{\mathbb{Q}\widetilde{\mathbb{Q}}}(t, \tau_1,\tau_2)\right)\ell^{\mathbb{Q}}(dz)$ for $t\in[0,\tau_1]$.
\end{Proposition}
\begin{proof}
	We know by definition that $\Pi_1^{\mathbb{Q}\widetilde{\mathbb{Q}}}$ is a continuous adapted process that is square-integrable and $\Pi_2^{\mathbb{Q}\widetilde{\mathbb{Q}}}$ is deterministic and càdlàg in time.
	Hence, all processes are predictable.
	Following \cite{OksendalSulem} (cf.\ Theorem 1.35), we need to show that \Eqref{eq:MG_Property} is satisfied, so that
	$Z^{\mathbb{Q}\widetilde{\mathbb{Q}}}$ is a true martingale.
	Considering the dynamics of $Z^{\mathbb{Q}\widetilde{\mathbb{Q}}}$ using Itô's formula, we have
	\begin{align*}
	d	Z^{\mathbb{Q}\widetilde{\mathbb{Q}}}(t,\tau_1,\tau_2)= 	Z^{\mathbb{Q}\widetilde{\mathbb{Q}}}(t-,\tau_1,\tau_2) \left[-\Pi_1^{\mathbb{Q}\widetilde{\mathbb{Q}}}(t, \tau_1,\tau_2)dW^{\mathbb{Q}}_t-\int_{\mathbb{R}}\Pi_2^{\mathbb{Q}\widetilde{\mathbb{Q}}}(t, \tau_1,\tau_2)z\widetilde{N}^{\mathbb{Q}}(dt, dz)\right]\;,
	\end{align*}
	so that $Z^{\mathbb{Q}\widetilde{\mathbb{Q}}}$ is a local $\mathbb{Q}$-martingale, where $W^{\mathbb{Q}}$ and $\widetilde{N}^{\mathbb{Q}}(\cdot,\cdot)$ are independent of each other.
	Hence, it is enough to show, that $Z_1^{\mathbb{Q}\widetilde{\mathbb{Q}}}$  and $Z_2^{\mathbb{Q}\widetilde{\mathbb{Q}}}$ are true martingales.
	We can prove Novikov's condition regarding the continuous part (see, e.g., \cite{Protter2005}, cf.\ Theorem 41, Chapter III.8) as  $\Pi_1^{\mathbb{Q}\widetilde{\mathbb{Q}}}$:
	\begin{align*}
	\mathbb{E}_{\mathbb{Q}}\left[e^{\frac{1}{2}\int_{0}^{\tau_1}\Pi_1^{\mathbb{Q}\widetilde{\mathbb{Q}}}(s,\tau_1,\tau_2)dW^{\mathbb{Q}}_s}\right]
	=e^{\frac{1}{2}\int_{0}^{\tau_1}\Pi_1^{\mathbb{Q}\widetilde{\mathbb{Q}}}(s,\tau_1,\tau_2)^2ds}<\infty\;.
	\end{align*}
	Hence, $Z_1^{\mathbb{Q}\widetilde{\mathbb{Q}}}$ is a true martingale.
	Moreover, $Z_2^{\mathbb{Q}\widetilde{\mathbb{Q}}}$ is a true martingale under $\mathbb{Q}$ since
	\begin{align*}
	\mathbb{E}_{\mathbb{Q}}[Z_2^{\mathbb{Q}\widetilde{\mathbb{Q}}}(\tau_1,\tau_1,\tau_2)]=
	\mathbb{E}_{\mathbb{Q}}\left[e^{\lambda^{\mathbb{Q}}\int_{0}^{\tau_1}\ln(1-\Pi_2^{\mathbb{Q}\widetilde{\mathbb{Q}}}(s, \tau_1,\tau_2))\widetilde{N}^{\mathbb{Q}}(ds,dz)}\right]
	e^{\int_{0}^{\tau_1}\int_{\mathbb{R}}\left(\ln(1-\Pi_2^{\mathbb{Q}\widetilde{\mathbb{Q}}}(s, \tau_1,\tau_2))+\Pi_2^{\mathbb{Q}\widetilde{\mathbb{Q}}}(s, \tau_1,\tau_2)\right)ds}=1\;,
	\end{align*}
	where the last equality follows from the Lévy-Khintchine representation and  \Cref{ass:TechnicalRequirementsQ3} in \Cref{app:tech_requ3}.
	Hence, we can apply Girsanov's Theorem (see, e.g., \cite{OksendalSulem}, cf.\ Theorem 1.35) and the assertion follows. \qed\\
\end{proof}
Note that the MPDP of diffusion and jump risk, $\Pi_1^{\mathbb{Q}\widetilde{\mathbb{Q}}}$ and $\Pi_2^{\mathbb{Q}\widetilde{\mathbb{Q}}}$, are negative following from Jensen's inequality.
Hence, the geometric averaging technique induces less risk than the application of the approximated arithmetic average for which we need to pay a cost of approximation risk.

\paragraph{On the Pricing Spread.}
We would like to compare the approximated swap price $F^a$ under $\mathbb{Q}$ with the swap price $F$ under $\widetilde{\mathbb{Q}}$. The diffusion part of the swap price dynamics coincides since we consider a deterministic volatility structure. The only differences are located in the compensator of the compound compensated Poisson process and the jump coefficient. If the jump coefficient is independent of delivery time, the distribution of $F^a$ under $\mathbb{Q}$ and the distribution of $F$ under $\widetilde{\mathbb{Q}}$ are the same. 
For differences in a stochastic volatility setting, we refer to \cite{Kemper2022}. 
For the swap prices $F$ and $F^a$, both under the artificial measure $\mathbb{Q}$, we have the following result:
\begin{Corollary} \label{lem:differenceAveraging}
	\begin{itemize}
	\item[$(i)$] The swap price $F$ is always smaller or equal than $F^A$.
	\item[$(ii)$]  
	The pricing spread between $F$ and $F^a$ is attained by
	\begin{align}
	&F^a(t,\tau_1,\tau_2)-F(t,\tau_1,\tau_2)= F^a(t,\tau_1,\tau_2)\Big[1-D(t,\tau_1,\tau_2)\Big]\;,\\
	&D(t,\tau_1,\tau_2)=e^{-\frac12 \int_{0}^{t} \mathbb{V} [\sigma(s,U)]ds-\int_{0}^{t}\int_{\mathbb{R}}\left(\ln \mathbb{E} [e^{\eta(s,U)z}]-\mathbb{E} [\eta(s,U)]z\right)N(ds, dz)}\;.
	\label{eq:numeraire}
	\end{align}
 \item[$(iii)$] If the jump coefficient, $\eta(t,u)$, is independent of the delivery time, then the swap price $F$ is smaller or equal than $F^a$. 
\end{itemize}
\end{Corollary}
\begin{proof}
	\begin{itemize}
		\item[$(i)$]
	The continuous arithmetic weighted average is greater than the geometric one, which directly follows from Jensen's inequality.
	\item[$(ii)$] 
	Using \Eqref{eq:arithAverage_drift}, we find that
	$
	F^a(t,\tau_1,\tau_2)=e^{\bar{X}^a(t,\tau_1,\tau_2)}
	$,
	where $\bar{X}^a(t,\tau_1,\tau_2)$ is the solution of the following arithmetic Brownian motion:
	\begin{align*}
	d\bar{X}^a(t,\tau_1,\tau_2)=&~\mathbb{E}[\sigma(t,U)]dW^\mathbb{Q}_t+\int_{\mathbb{R}}\ln \mathbb{E} [e^{\eta(t,U)z}]\widetilde{N}^{\mathbb{Q}}(dt, dz)\\&~-\left(\frac12\mathbb{E}[\sigma(t,U)]^2+\int_{\mathbb{R}} \mathbb{E} [e^{\eta(t,U)z}]-1-\ln\mathbb{E}[e^{\eta(t,U)z}]\right)dt\;.
	\end{align*}
	Taking \Cref{eq:Xbar} into account, the pricing spread  is given by
	\begin{align*}
	F^a(t,\tau_1,\tau_2)-F(t,\tau_1,\tau_2)=F^a(t,\tau_1,\tau_2)\left(1-D(t,\tau_1,\tau_2)\right)\;,
		\end{align*}
		such that 
		$
		F(t,\tau_1,\tau_2)=F^a(t,\tau_1,\tau_2)D(t,\tau_1,\tau_2)$,
		where
		\begin{align*}
	D(t,\tau_1,\tau_2)=e^{\bar{X}(t,\tau_1,\tau_2)-\bar{X}^a(t,\tau_1,\tau_2)}  =e^{-\frac12 \int_{0}^{t} \mathbb{V} [\sigma(s,U)]ds-\int_{0}^{t}\int_{\mathbb{R}}\ln \mathbb{E} [e^{\eta(s,U)z}]-\mathbb{E} [\eta(s,U)]zN(ds, dz)}\;.
	\end{align*}
 \item[$(iii)$] If $\eta(t,u)\independent u$, then 
 \begin{align*}
	D(t,\tau_1,\tau_2)=e^{-\frac12 \int_{0}^{t} \mathbb{V} [\sigma(s,U)]ds}\;.
	\end{align*}
	Since $\mathbb{V} [\sigma(\cdot,U)]\geq 0$ by Jensen, it follows that $D(t,\tau_1,\tau_2)\in (0,1]$ and thus $F\leq F^a$.	\qed
\end{itemize}
\end{proof}

We conclude that arithmetic and in specific cases approximated averaging lead to higher swap prices than the geometric average. 
We would like to stress that $D$ in \Eqref{eq:numeraire} is not affected by measure changes since it is characterized by a drift component and a pure jump component exclusively (see also \Eqref{eq:Numeraire}).
Moreover, note that $D$ can be seen as stochastic discount factor, which can be used to derive the swap price $F$ given $F^a$.
Vice versa, consider
	\begin{align}
	F^a(t,\tau_1,\tau_2)=F(t,\tau_1,\tau_2) D^{-1}(t,\tau_1,\tau_2)\;.
	\end{align}
	The exponential part of $ D^{-1}$ can be interpreted as a price (premium) per share, which we pay for an imprecise averaged swap.
	Moreover, we can see $D$ as the price process of a non-dividend paying asset evolving as
	\begin{align}
	\frac{dD(t,\tau_1,\tau_2)}{D(t-,\tau_1,\tau_2)} = -\frac12 \mathbb{V}[\sigma(t,U)]dt+\int_{\mathbb{R}}\left(\frac{e^{\mathbb{E}[\eta(t,U)]z}}{\mathbb{E}[e^{\eta(t,U)z}]}-1\right)N(dt,dz)\;, \label{eq:Numeraire}
	\end{align}
	such that we can interpret $D$ as a numéraire. 
	If $F^a$ is a martingale, then $\frac{F}{D}$ is also a martingale.
	If $F$ is a martingale, then $F^aD$ is a martingale (see, e.g., \cite{Shreve2004}, cf.\ Theorem 9.2.2).
	We can thus use it to price options and other derivatives on the swap. In the subsequent section, we introduce the model under its physical measure $\mathbb{P}$.

\begin{Remark}[Delivery-Dependent Intensity]\label{app:DeliveryDependentIntensity}
Let us consider an adjusted version of the artificial futures price under the artificial pricing measure $\mathbb{Q}$ similar to \Eqref{eq:futuresQ} given by
\begin{align}
\frac{df(t,\tau)}{f(t-,\tau)}= \sigma(t,\tau)dW^{\mathbb{Q}}_t+\int_{\mathbb{R}}\left(e^{\eta z}-1\right)\widetilde{N}^\tau(dt, dz)\;,
\end{align}
where $\eta\in\mathbb{R}$ and the compensated Poisson random measure is defined by $\widetilde{N}^\tau(dt, dz):={N}(dt, dz)-\lambda^\mathbb{Q}(\tau)G(dz)dt$,
with a jump intensity  adjusted to a deterministic, positive, and bounded function of the \textit{delivery time}.
\begin{itemize}
\item[$(i)$] Analogous to \Cref{lemma:dyn_Swap_Q},  the dynamics of the swap price, defined by geometric averaging, are given  by
	\begin{equation}
	\begin{split}
	\frac{d F(t,\tau_1,\tau_2)}{F(t-,\tau_1,\tau_2)} = &~
	\mathbb{E}[\sigma(t,U)]~dW^{\mathbb{Q}}_t+\int_{\mathbb{R}} \left(e^{\eta z}-1\right){N}(dt,dz)\\
	&~-\left(\frac{1}{2}\mathbb{V}\left[\sigma(t,U) \right]
	+ \mathbb{E}[\lambda^\mathbb{Q}(U)]\int_{\mathbb{R}} \left(e^{\eta z}-1\right)G(dz) \right) dt\;, \label{eq:NewSwap}
	\end{split}
	\end{equation}
	where $U$is the random delivery variable with density $w(u,\tau_1,\tau_2)$.
	\item[$(ii)$] If the volatility is independent of delivery time, then the approximated and geometric average coincide and so their risk-neutral pricing measure.
	However, the resulting swap price process in \Eqref{eq:NewSwap} is not a martingale under $\mathbb{Q}$ since the intensity is affected by the averaging procedure. 
	\item[$(iii)$] In the case of delivery-dependent volatility, the MPDP from \Cref{Def_MPDP} adjusts to  
	$(\Pi_1^{\mathbb{Q}\widetilde{\mathbb{Q}}},0)$. Hence, the MPDP associated to the Brownian motion stays the same  and its second dimension becomes zero since the jump coefficient $\eta$ is independent of delivery time. However, under the assumption that the intensity is delivery-dependent, the swap price is in general not a martingale under~$\widetilde{\mathbb{Q}}$.
	\item[$(iv)$] The swap price is a $\widetilde{\mathbb{Q}}-$martingale, only if the swap's jump intensity under $\widetilde{\mathbb{Q}}$ is given by $\mathbb{E}[\lambda^\mathbb{Q}(U)]$, i.e., if
	$
	\widetilde{N}^{\tau_1,\tau_2}(dt, dz):={N}(dt, dz)-\mathbb{E}[\lambda^\mathbb{Q}(U)]G(dz)dt
	$,
	is a compensated Poisson random measure under $\widetilde{\mathbb{Q}}$.
	\end{itemize}
\end{Remark}

\section{The Real-World Model} \label{sec:Model}
A typical feature of electricity prices beyond seasonalities  and the Samuelson effect is the mean-reverting behavior (see, e.g., \cite{Benth2008} and \cite{Benth2019}).
In order to implement the drift feature, we derive the futures under the physical measure $\mathbb{P}$. Note that we will include mean-reversion at the futures and thus the swap's \textit{rate} level.
We then consider the resulting market prices of risk transferring to the artificial and the swap's risk-neutral measure.

\subsection{The Swap Price under the Physical Measure}\label{sec:PhysicalMgMeasure}
We now derive the price of a swap contract that delivers one unit of electricity during the fixed delivery period $(\tau_1,\tau_2]$, similar to \Cref{sec:AVERAGING} but now under the physical measure $\mathbb{P}$.
Hence, starting from the physical measure~$\mathbb{P}$,  the logarithmic futures price process from \Eqref{eq:futuresQsol1}, given by
\begin{equation}
\begin{split}
\ln f(t,\tau) =~& e^{-\int_{0}^{t}\kappa(s)ds}\ln f(0,\tau) + \int_{0}^{t}e^{-\int_{v}^{t}\kappa(q)dq}\mu(v,\tau)dv\\&
		+\int_{0}^{t}e^{-\int_{v}^{t}\kappa(q)dq}\sigma(v,\tau)dW^{\mathbb{P}}_v +\int_{0}^{t}e^{-\int_{v}^{t}\kappa(q)dq}\eta(v,\tau)d\widetilde{J}^{\mathbb{P}}_v\;, \label{eq:LogFuturesP}
\end{split}
\end{equation}
where $W^{\mathbb{P}}$ is a Brownian motion under the physical measure $\mathbb{P}$ independent of the compound compensated jump process $\widetilde{J}^{\mathbb{P}}$. In particular, $\widetilde{J}^{\mathbb{P}}$ is defined through the $\mathbb{P}$-compensated Poisson random measure $\widetilde{N}^{\mathbb{P}}(dt,dz)=N(dt,dz)-\ell^{\mathbb{P}}(dz)dt$ with Lévy measure $\ell^{\mathbb{P}}(dz) = \lambda^{\mathbb{P}}G(dz)$ that is independent of delivery time. Note that $\lambda^{\mathbb{P}}>0$ indicates the jump intensity under the physical measure and $G(dz)$ is the jump size distribution.\\

In order to characterize the futures price  in more detail, we introduce the following lemma.
\begin{Lemma}
	We assume that the coefficients satisfy suitable integrability and measurability conditions (see \Cref{ass:TechnicalRequirementsP} in \Cref{app:tech_requ3})
	such that 
\Cref{eq:LogFuturesP} is the unique strong solution to the dynamics
\begin{equation}
\begin{split}
    \frac{df(t,\tau)}{f(t-,\tau)}=  \sigma(t,\tau)dW^{\mathbb{P}}_t + \int_{\mathbb{R}}\left(e^{\eta(t,\tau)z}-1\right)\widetilde{N}^{\mathbb{P}}(dt,dz)+c^{\mathbb{P}}(t,\tau,\ln f(t,\tau))dt\;, \label{eq:dynFuturesP}
\end{split}
\end{equation}
where the drift-term is characterized by
\begin{align}
c^{\mathbb{P}}(t,\tau,Y) =\mu(t,\tau)  - \kappa(t)Y+ \frac12\sigma(t,\tau)^2 + \psi^{\mathbb{P}}(\eta(t,\tau))\;.
\end{align}
Hence, the logarithmic futures evolves as 
\begin{align}
d\ln f(t,\tau)=\left(\mu(t,\tau)-\kappa(t)\ln f(t,\tau)\right)dt+ \sigma(t,\tau)dW^{\mathbb{P}}_t + \eta(t,\tau) d\widetilde{J}^{\mathbb{P}}_t\;.\label{eq:AdjOU2}
\end{align}
\end{Lemma}
\begin{proof}
	The unique strong solution follows from \cite{Benth2008} (cf.\ Proposition 3.1).
Applying Ito's formula leads to the desired dynamics (see \cite{OksendalSulem}, cf.\ Theorem 1.16).\qed\\
\end{proof}

Note that the assumption behind the model induces a finite second moment as well as a finite moment generating function of the jump size distribution. In \Cref{ex:JumpsNormal,ex:JumpsExponential}, we consider suitable distributions for these jump sizes.

\begin{Remark} \label{re:lognormal}
In the literature, we sometimes find the application of lognormal distributed jump sizes (see, e.g., \cite{BorovkovaPermana2006} and \cite{BorovkovaSchmeck2017}).
This distribution, however, is not suitable for our setting since its moment generating function $\mathbb{E}[e^{\eta Z}]$ is not finite at any positive value $\eta$ (see, e.g., \cite{GrayPitts2012}, cf.\ Chapter 2.2.6). 
Hence, the lognormal distribution contradicts the integrability assumption  in \Cref{ass:TechnicalRequirementsP}~$(i)$ under the physical measure in \Cref{app:tech_requ3}.
\end{Remark}

As in the previous section, we now derive the swap prices  resulting from geometric and approximated averaging.

\begin{Lemma}[The Swap Price under $\mathbb{P}$] \label{lem:SwapPGeom}
Let \Cref{ass:TechnicalRequirementsP,ass:StochasticFubini} in  \Cref{app:tech_requ3} be satisfied.
Then, 
the swap price based on geometric averaging evolves as
\begin{equation}
\begin{split}
\frac{dF(t,\tau_1,\tau_2)}{F(t-,\tau_1,\tau_2)} =&~
\mathbb{E}[\sigma(t,U)]dW^{\mathbb{P}}_t + \int_{\mathbb{R}}\left(e^{\mathbb{E}[\eta(t,U)]z}-1\right)  \widetilde{N}^{\mathbb{P}}(dt,dz) + \widetilde{c}^{\mathbb{P}}(t,\tau_1,\tau_2,\ln F(t,\tau_1,\tau_2))dt\;, \label{eq:SwapEvolP}
\end{split}
\end{equation}
where the drift term is given by
\begin{align}
\widetilde{c}^{\mathbb{P}}(t,\tau_1,\tau_2,\bar{Y})=\mathbb{E}[\mu(t,U)]-\kappa(t)\bar{Y} + \frac12\mathbb{E}[\sigma(t,U)]^2 + \psi^{\mathbb{P}}(\mathbb{E}[\eta(t,U)])\;.
\end{align}
\end{Lemma}
\begin{proof}
Following the considerations in the previous section, the swap price is defined by the geometric average in \Eqref{eq:GeomAverage}. Using the integral representation of \Cref{eq:AdjOU2} and the stochastic Fubini theorem (see \cite{Protter2005}, cf.\ Theorem 65), we can introduce the dynamics of the swap's logarithmic return by
\begin{equation}
\begin{split}
d\ln F(t,\tau_1,\tau_2) =&~ \left(\mathbb{E}[\mu(t,U)]-\kappa(t)\ln F(t,\tau_1,\tau_2)\right)dt  + \mathbb{E}[\sigma(t,U)]dW^{\mathbb{P}}_t+\mathbb{E}[\eta(t,U)]d\widetilde{J}^{\mathbb{P}}_t\;.\label{eq:Y2tilde}
\end{split}
\end{equation} 
An application of Ito's formula (see \cite{OksendalSulem}, cf.\ Theorem 1.16) yields the desired swap dynamics. \qed\\
\end{proof}

Note that the speed of mean-reversion $\kappa(t)$ has to be independent of the delivery time. This assumption ensures that $\ln F$, in \Eqref{eq:Y2tilde}, is again an Ornstein-Uhlenbeck process and that the swap's price dynamics in \Eqref{eq:SwapEvolP} stay tractable. This is also in line with the findings in \cite{Benth2019} (cf.~Proposition 2.2) and \cite{Latini2019}.
In particular, the mean-reverting effect comprises the jump component as well, even if we implement it through a measure change of the Brownian part. 
More precisely, mean-reversion connected to jumps covers indeed a unique feature of electricity markets known as spikes: Spikes are large jumps quickly returning to the ``normal'' level (see, e.g., \cite{Klüppelberg2010}). They arise as electricity is not storable on a large scale and since the electricity demand is not elastic (see \cite{BorovkovaSchmeck2017}).\\

Let us now investigate the swap price under the physical measure resulting from approximated averaging (see \Cref{eq:arithAverage_drift} in order to compare the pricing spread between both approaches.
\begin{Lemma} \label{lem:SwapPArithm}
Let   \Cref{ass:TechnicalRequirementsP,ass:StochasticFubini} in \Cref{app:tech_requ3} be satisfied.
Then, the swap price dynamics based on \textit{approximated averaging} evolve as
\begin{equation}
\begin{split}
\frac{dF^a(t,\tau_1,\tau_2)}{F^a(t-,\tau_1,\tau_2)} =
\mathbb{E}[\sigma(t,U)]dW^{\mathbb{P}}_t + \int_{\mathbb{R}}\left(\mathbb{E}[e^{\eta(t,U)z}]-1\right)  \widetilde{N}^{\mathbb{P}}(dt,dz) +\mathbb{E}_U[c^{\mathbb{P}}(t,U,\ln f(t,U))]dt\;, \label{eq:DynArithmSwapP}
\end{split}
\end{equation}
with $\mathbb{E}_U$ denoting the expectation with respect to the random delivery variable $U$ having density $w(u,\tau_1,\tau_2)$.
\end{Lemma}
\begin{proof}
We use the approximated averaging methodology (see \Eqref{eq:arithAverage_drift}) in order to derive the swap price evolution and apply the stochastic Fubini theorem (see \cite{Protter2005}, cf.\ Theorem 65) leading to \Eqref{eq:DynArithmSwapP}. 
\qed\\
\end{proof}

Similar to \Cref{lem:differenceAveraging}, we now evaluate the pricing spread between the exact geometric averaged swap resulting from \Cref{lem:SwapPGeom} and the approximated version resulting from \Cref{lem:SwapPArithm} under the physical measure in the next corollary.
\begin{Corollary}
The spread between the swap prices $F$ and $F^a$ under $\mathbb{P}$ coincides with the pricing spread from \Cref{lem:differenceAveraging}~(ii).
\end{Corollary}
As the numéraire in \Eqref{eq:Numeraire} is not affected by a change of measure, the pricing spread between the exact and approximated swap price stays the same independent of the measure.

\subsection{The Swap Price $F$ under its Risk-Neutral Measure $\widetilde{\mathbb{Q}}$} \label{sec:MeasureChangeF}
In order to derive the swap's martingale measure $\widetilde{\mathbb{Q}}$, we introduce the \textit{true} market price of risk for the swap price resulting from geometric averaging in the next definition:
\begin{Definition} \label{def:MarketPricesGeom}
		We define the true market price of risk for the swap by $\Pi^{\mathbb{P}\widetilde{\mathbb{Q}}}:=(\Pi_1^{\mathbb{P}\widetilde{\mathbb{Q}}},\Pi_2^{\mathbb{P}\widetilde{\mathbb{Q}}})$,  where
		\begin{align}
		\Pi_1^{\mathbb{P}\widetilde{\mathbb{Q}}}(t,\tau_1,\tau_2)&:=
		\frac{\mathbb{E}[\mu(t,U)]-\kappa(t)\ln F(t,\tau_1,\tau_2)+\frac12\mathbb{E}[\sigma(t,U)]^2 }{\mathbb{E}[\sigma(t,U)]} \;, \label{eq:MPR_F_1}\\
		\Pi_2^{\mathbb{P}\widetilde{\mathbb{Q}}}(t,\tau_1,\tau_2)&:= 
		1-\int_{\mathbb{R}}z\ell^{\mathbb{P}}(dz)\frac{\mathbb{E}[\eta(t,U)]}{\int_{\mathbb{R}}\left(e^{\mathbb{E}[\eta(t,U)]z}-1\right)\ell^{\mathbb{P}}(dz)}\;.\label{eq:MPR_F_2}
		\end{align}
\end{Definition}

Note that the market price of risk does not enter the jump size distribution since we restrict $\Pi_2^{\mathbb{P}\widetilde{\mathbb{Q}}}$ to depend on trading time and delivery period. Hence, the market price of jump risk affects the jump intensity only.\\

We follow the methodology of \cite{Benth2019} to change the measure from the physical measure $\mathbb{P}$ to the swap's risk-neutral measure $\widetilde{\mathbb{Q}}$. Therefore, let $\pi=(\pi_1,\pi_2)$ be a predictable process satisfying 
\begin{align}
\mathbb{E}\left[\int_{0}^{\tau_1}\norm{\pi(s,\tau_1,\tau_2)}^2ds\right]<\infty\;. \label{eq:piSquareIntegrable}
\end{align}
We define a new process $Z^{\mathbb{P}\widetilde{\mathbb{Q}}}$ being the unique strong solution of 
\begin{align}
dZ^{\mathbb{P}\widetilde{\mathbb{Q}}}(t,\tau_1,\tau_2)=Z^{\mathbb{P}\widetilde{\mathbb{Q}}}(t-,\tau_1,\tau_2)dH(t,\tau_1,\tau_2)\;, \label{eq:densitydynamics}
\end{align}
such that $Z^{\mathbb{P}\widetilde{\mathbb{Q}}}(0,\tau_1,\tau_2)=1$, where 
\begin{align}
dH(t,\tau_1,\tau_2) = \pi_1(t,\tau_1,\tau_2)dW^{\mathbb{P}}_t+\pi_2(t,\tau_1,\tau_2)d\widetilde{J}^{\mathbb{P}}_t\;. \label{eq:H}
\end{align}
If $\pi_j$ satisfies \Eqref{eq:piSquareIntegrable}, then $H$ is a well-defined square integrable martingale. Note that the process $Z^{\mathbb{P}\widetilde{\mathbb{Q}}}$ is known as the Doléans-Dade exponential of $H$ that is explicitly given by
\begin{align}
Z^{\mathbb{P}\widetilde{\mathbb{Q}}}(t,\tau_1,\tau_2) = e^{H(t,\tau_1,\tau_2)-\frac12\int_{0}^{t}\pi_1(s,\tau_1,\tau_2)^2 ds} \prod_{0<s\leq t}\left(1+\Delta H(s,\tau_1,\tau_2)\right)e^{-\Delta H(s,\tau_1,\tau_2)}\;. \label{eq:DoleanDadeExponential}
\end{align}
If $Z^{\mathbb{P}\widetilde{\mathbb{Q}}}$ is a strictly positive martingale, then we can define the equivalent probability measure $\widetilde{\mathbb{Q}}$ by
\begin{align}
\frac{d\widetilde{\mathbb{Q}}}{d \mathbb{P}} = Z^{\mathbb{P}\widetilde{\mathbb{Q}}}(\tau_1,\tau_1,\tau_2)\;,
\end{align} 
where $Z^{\mathbb{P}\widetilde{\mathbb{Q}}}$ functions as the Radon-Nikodym derivative. If we further assume that $\mathbb{E}_{\mathbb{P}}[Z^{\mathbb{P}\widetilde{\mathbb{Q}}}(\tau_1,\tau_1,\tau_2)]=1$, then Girsanov's theorem (see \cite{OksendalSulem}, cf.\ Theorem 1.35) states for $\pi:=-\Pi^{\mathbb{P}\widetilde{\mathbb{Q}}}$ that 
\begin{align}
W^{\widetilde{\mathbb{Q}}}_t = W^{\mathbb{P}}_t + \int_{0}^{t} \Pi_1^{\mathbb{P}\widetilde{\mathbb{Q}}}(s,\tau_1,\tau_2)ds\;,
\end{align}
is a Brownian motion with respect to $\widetilde{\mathbb{Q}}$ and
\begin{align}
\widetilde{N}^{\widetilde{\mathbb{Q}}}(dt,dz) =\widetilde{N}^{\mathbb{P}}(dt,dz) + \Pi_2^{\mathbb{P}\widetilde{\mathbb{Q}}}(t,\tau_1,\tau_2)\ell^{\mathbb{P}}(dz)dt\;,
\end{align}
is a $\widetilde{\mathbb{Q}}$-compensated Poisson random measure of $N(\cdot,\cdot)$.\\

Under the above assumptions specified later a straightforward valuation leads to the following result:
\begin{Proposition} \label{prop:MeasureChangeF}
The swap price process $F$ defined in \Eqref{eq:GeomAverage} is a martingale under $\widetilde{\mathbb{Q}}$ given by
\begin{align}
\frac{dF(t,\tau_1,\tau_2)}{F(t-,\tau_1,\tau_2)} =
\mathbb{E}[\sigma(t,U)]dW^{\widetilde{\mathbb{Q}}}_t + \int_{\mathbb{R}}\left(e^{\mathbb{E}[\eta(t,U)]z}-1\right)  \widetilde{N}^{\widetilde{\mathbb{Q}}}(dt,dz)\;.
\end{align}
\end{Proposition}

We would like to investigate the consequences of our previous assumptions.
\begin{Remark}
	\begin{itemize}
	\item[$(i)$] The Doléans-Dade exponential in \Eqref{eq:DoleanDadeExponential} is positive if $\pi_2(s-)\Delta J>-1$, i.e., if $\Pi_2^{\mathbb{P}\widetilde{\mathbb{Q}}}\Delta J<1$. Hence, similar to \cite{Benth2019}, we need to assume that the market price of jump risk is bounded and deterministic over the entire time period such that $\Pi_2^{\mathbb{P}\widetilde{\mathbb{Q}}}(t,\tau_1,\tau_2)z<1$ for $\ell^{\mathbb{P}}$-a.e.\ $z\in\mathbb{R}$ and for each $t\in[0,\tau_1]$.
\item[$(ii)$]
If $\ln f$, and so $\ln F$, is driven by a compensated Poisson process only,
then the swap's market price of risk is attained by $\Pi^{\mathbb{P}\widetilde{\mathbb{Q}}}:=(0,\Pi_2^{\mathbb{P}\widetilde{\mathbb{Q}}})$, where
\begin{align}
\Pi_2^{\mathbb{P}\widetilde{\mathbb{Q}}}(t,\tau_1,\tau_2):= 
1-\frac{\mathbb{E}[\eta(t,U)]\int_{\mathbb{R}}z\ell^{\mathbb{P}}(dz)}{\int_{\mathbb{R}}\left(e^{\mathbb{E}[\eta(t,U)]z}-1\right)\ell^{\mathbb{P}}(dz)} + 
\frac{\mathbb{E}[\mu(t,U)]-\kappa(t)\ln F(t,\tau_1,\tau_2)}{\int_{\mathbb{R}}\left(e^{\mathbb{E}[\eta(t,U)]z}-1\right)\ell^{\mathbb{P}}(dz)}\;.
\end{align}
In this setting, we need to require that $\kappa(t)\equiv 0$.
	\end{itemize}
\end{Remark}

Note that a positive local martingale is a supermartingale. 
Hence, in order to prove that the Radon-Nikodym density $Z^{\mathbb{P}\widetilde{\mathbb{Q}}}$ is a true martingale, it is sufficient to verify that $\mathbb{E}_{\mathbb{P}}[Z^{\mathbb{P}\widetilde{\mathbb{Q}}}(\tau_1,\tau_1,\tau_2)]=1$ is satisfied, which is proven in the next proposition. 
\begin{Proposition} \label{prop:ZisMartingale}
	Under \Cref{ass:TechnicalRequirementsP3} in \Cref{app:tech_requ3},
	the process $Z^{\mathbb{P}\widetilde{\mathbb{Q}}}$ defined by \Eqref{eq:densitydynamics} is a strictly positive true martingale.
\end{Proposition}
\begin{proof} 
	In \Cref{app:Proof_MeasureChangeF}, we prove this proposition  even in a stochastic volatility framework. \qed\\
\end{proof}

\subsection{The Approximated Swap Price $F^a$ under the Artificial Risk-Neutral Measure}
We introduce the \textit{classical} market price of risk for the approximated swap price in the next definition.
\begin{Definition} \label{def:MarketPricesAppr}
We define the classical market price of risk for the approximated swap by $\Pi^{\mathbb{P}\mathbb{Q}}:=(\Pi_1^{\mathbb{P}\mathbb{Q}},\Pi_2^{\mathbb{P}\mathbb{Q}})$,  where
\begin{align}
\Pi_1^{\mathbb{P}\mathbb{Q}}(t,\tau_1,\tau_2)&:=
\frac{\mathbb{E}[\mu(t,U)]-\kappa(t)\ln F(t,\tau_1,\tau_2)+\frac12\mathbb{E}[\sigma^2(t,U)] }{\mathbb{E}[\sigma(t,U)]} \;,\\
\Pi_2^{\mathbb{P}\mathbb{Q}}(t,\tau_1,\tau_2)&:= 
1-\int_{\mathbb{R}}z\ell^{\mathbb{P}}(dz)\frac{\mathbb{E}[\eta(t,U)]}{\int_{\mathbb{R}}\left(\mathbb{E}[e^{\eta(t,U)z}]-1\right)\ell^{\mathbb{P}}(dz)}\;.
\end{align}
\end{Definition}

Note that we assume that the market price of jump risk affects the jump intensity only. The market price of risk does not enter the jump size distribution since we restrict $\Pi_2^{\mathbb{P}\mathbb{Q}}$ to depend on trading and delivery period.\\

Similar to the last subsection, we can define the equivalent (artificial) probability measure $\mathbb{Q}$ by
\begin{align}
\frac{d\mathbb{Q}}{d \mathbb{P}} = Z^{\mathbb{P}\mathbb{Q}}(\tau_1,\tau_1,\tau_2)\;,
\end{align} 
where $Z^{\mathbb{P}\mathbb{Q}}$ functions as the Radon-Nikodym derivative characterized by $\pi:=-\Pi^{\mathbb{P}\mathbb{Q}}$. If we further assume that $\mathbb{E}_{\mathbb{P}}[Z^{\mathbb{P}\mathbb{Q}}(\tau_1,\tau_1,\tau_2)]=1$, then Girsanov's theorem (see \cite{OksendalSulem}, cf.\ Theorem 1.35) states that
\begin{align}
W^{\mathbb{Q}}_t = W^{\mathbb{P}}_t + \int_{0}^{t} \Pi_1^{\mathbb{P}\mathbb{Q}}(s,\tau_1,\tau_2)ds\;, 
\end{align}
is a Brownian motion with respect to $\mathbb{Q}$ and
\begin{align}
\widetilde{N}^{\mathbb{Q}}(dt,dz) =\widetilde{N}^{\mathbb{P}}(dt,dz) + \Pi_2^{\mathbb{P}\mathbb{Q}}(t,\tau_1,\tau_2)\ell^{\mathbb{P}}(dz)dt\;,
\end{align}
is a $\mathbb{Q}$-compensated Poisson random measure of $N(\cdot,\cdot)$.\\

Under the above assumptions a straightforward valuation leads to the following result:
\begin{Proposition}
	The approximated swap price process $F^a$ defined in \Eqref{eq:arithAverage_drift} is a martingale under $\mathbb{Q}$ given by
	\begin{align}
	\frac{dF^a(t,\tau_1,\tau_2)}{F^a(t-,\tau_1,\tau_2)} =
	 \mathbb{E}[\sigma(t,U)]dW^{\mathbb{Q}}_t + \int_{\mathbb{R}}\left(\mathbb{E}[e^{\eta(t,U)z}]-1\right)  \widetilde{N}^{\mathbb{Q}}(dt,dz)\;.
	\end{align}
\end{Proposition}
We refer to \Cref{sec:MeasureChangeF} for the consequences of the assumptions made above.

\subsection{The Decomposition of the Market Price of Risk} \label{sec:DecompositionofMP}
From the previous subsections, we know the corresponding market prices of risk for the swap price resulting from geometric averaging (see \Cref{def:MarketPricesGeom}) and from approximated averaging (see \Cref{def:MarketPricesAppr}). 
In this subsection, we identify a clear distinction between both market prices of risk leading to a specific decomposition that is strongly connected to the MPDP.\\

We now introduce the \textit{decomposition} of the true market price of risk, from \Cref{def:MarketPricesGeom}, which finally connects the classical market price of risk, specified in \Cref{def:MarketPricesAppr}, and the MPDP, defined in \Cref{Def_MPDP}.
The decomposition result is stated in the next proposition.
\begin{Proposition} \label{prop:Spread}
	The swap's true market price of risk, $\Pi^{\mathbb{P}\widetilde{\mathbb{Q}}}$,  resulting from geometric averaging (see \Cref{def:MarketPricesGeom}), decomposes into
	\begin{align}
	\Pi_j^{\mathbb{P}\widetilde{\mathbb{Q}}}(t,\tau_1,\tau_2)=
	\Pi_j^{\mathbb{P}\mathbb{Q}}(t,\tau_1,\tau_2) + \bar{\Pi}_j^{\mathbb{Q}\widetilde{\mathbb{Q}}}(t,\tau_1,\tau_2)\;,  \quad \text{for } j=1,2\;,
	\end{align}
	where $\Pi_j^{\mathbb{P}\mathbb{Q}}$ is specified in \Cref{def:MarketPricesAppr} and $\bar{\Pi}_j^{\mathbb{Q}\widetilde{\mathbb{Q}}}$ defines the  spread of diffusion and jump risk. More precisely,
	\begin{align}
	\bar{\Pi}_1^{\mathbb{Q}\widetilde{\mathbb{Q}}}(t,\tau_1,\tau_2) &= -\frac12 \frac{\mathbb{V}[\sigma(t,U)]}{\mathbb{E}[\sigma(t,U)]}\;, \label{eq:Spread1} \\
	\bar{\Pi}_2^{\mathbb{Q}\widetilde{\mathbb{Q}}}(t,\tau_1,\tau_2) &= 
	-\mathbb{E}[\eta(t,U)]\int_{\mathbb{R}}zG(dz)\frac{\int_{\mathbb{R}}\mathbb{E}[e^{\eta(t,U)z}]-e^{\mathbb{E}[\eta(t,U)]z}G(dz) }{\int_{\mathbb{R}}\left(\mathbb{E}[e^{\eta(t,U)z}]-1\right)G(dz)\int_{\mathbb{R}}\left(e^{\mathbb{E}[\eta(t,U)]z}-1\right)G(dz)}\;, \label{eq:Spread2}
	\end{align}
	where $\bar{\Pi}_2^{\mathbb{Q}\widetilde{\mathbb{Q}}}$ is independent of the jump intensity.
\end{Proposition}
\begin{proof}
The result is attained by subtracting the true market price of risk $\Pi^{\mathbb{P}\widetilde{\mathbb{Q}}}$ defined in \Cref{def:MarketPricesGeom} from the classical market price of risk $\Pi^{\mathbb{P}{\mathbb{Q}}}$ defined in \Cref{def:MarketPricesAppr}. \qed \\
\end{proof}

Hence, we found a representation of the true market price of risk of the swap price $F$, characterized by the classical market price of risk of the approximated swap $F^a$ and the spread $\bar{\Pi}^{\mathbb{Q}\widetilde{\mathbb{Q}}}=\left(\bar{\Pi}_1^{\mathbb{Q}\widetilde{\mathbb{Q}}},\bar{\Pi}_2^{\mathbb{Q}\widetilde{\mathbb{Q}}}\right)$.
We further investigate the spread in the next lemma.

\begin{Lemma}
	\begin{itemize}
		\item[$(i)$] The spread of diffusion risk, $\bar{\Pi}_1^{\mathbb{Q}\widetilde{\mathbb{Q}}}(t,\tau_1,\tau_2)$, is negative for all trading times $t\in [0,\tau_1]$.
		\item[$(ii)$] If the average jump size is positive, i.e., if $\int_{\mathbb{R}}zG(dz)>0$, then  the spread of jump risk, $\bar{\Pi}_2^{\mathbb{Q}\widetilde{\mathbb{Q}}}$, is negative.
		\item[$(iii)$] If the volatility is independent of the delivery, i.e., if $\sigma(t,u)\independent u$, then the spread of diffusion risk is zero, i.e., $\bar{\Pi}_1^{\mathbb{Q}\widetilde{\mathbb{Q}}}(t,\tau_1,\tau_2)\equiv 0$.
		\item[$(iv)$] If the jump coefficient is independent of the delivery, i.e., if $\eta(t,u)\independent u$, then the spread of jump risk is zero, i.e., $\bar{\Pi}_2^{\mathbb{Q}\widetilde{\mathbb{Q}}}(t,\tau_1,\tau_2)\equiv 0$.  	
	\end{itemize}
\end{Lemma}
\begin{proof}
The results in $(i)$ and $(ii)$ follow directly from Jensen's inequality.
The results in $(iii)$ and $(iv)$  follow from the fact that the numerator becomes zero whenever the delivery period disappears.  \qed\\
\end{proof}
As a result, whenever the spread $\bar{\Pi}_j^{\mathbb{Q}\widetilde{\mathbb{Q}}}$ is negative for $j=1,2$, then the approximated swap induces more risk than the swap price based on geometric averaging.
In particular, the considered spread has the same properties as the MPDP (see \cite{Kemper2022}).
Indeed, a comparison with our previous considerations in \Cref{sec:AVERAGING} gives the following insights:

\begin{Remark}
\begin{itemize}
	\item[$(i)$] The spread of diffusion risk, $\bar{\Pi}_1^{\mathbb{Q}\widetilde{\mathbb{Q}}}$, coincides with the MPDP of diffusion risk, $\Pi_1^{\mathbb{Q}\widetilde{\mathbb{Q}}}$, in \Eqref{eq:AddMP1} from \Cref{sec:AVERAGING}.
	\item[$(ii)$] The spread of jump risk, $\bar{\Pi}_2^{\mathbb{Q}\widetilde{\mathbb{Q}}}$, does not  coincide with the MPDP of jump risk, $\Pi_2^{\mathbb{Q}\widetilde{\mathbb{Q}}}$, from \Eqref{eq:AddMP2} but with $\Pi_2^{\mathbb{Q}\widetilde{\mathbb{Q}}}(1-\Pi_2^{\mathbb{P}\mathbb{Q}})$.
	This connection occurs naturally by the change of measure. 
\end{itemize}
\end{Remark}

Hence, starting from the physical measure, we can find the swaps true martingale measure based on the true market price of risk defined in \Cref{def:MarketPricesGeom}.
If we would like to adjust already existing models using the classical market price, we can easily adjust the model through the spread defined in \Cref{prop:Spread} that is strongly connected to the MPDP defined in \Cref{Def_MPDP}.

\section{Conclusion} \label{sec:summary}
In this paper, we provide analytical characteristics of the MPDP for electricity swaps embedded in a Merton type model.\\

We adjust the Heston type setting of \cite{Kemper2022} to a jump framework of Merton type leading to the MPDP for diffusion \textit{and} jump risk.
We thus identify the MPDP for the jump component, which turns out to be negative as it is the case for the diffusion component.
In addition, we transfer the model to the physical measure under which we allow for mean-reversion adn delivery-dependent effects such as seasonalities and term-structure effects.
A comparison of the risk-neutral measures of the swap resulting from geometric and approximated averaging, $\mathbb{Q}$ and $\widetilde{\mathbb{Q}}$, offers the decomposition of the ``true'' market price of risk comprising the ``classical'' market price of risk and the MPDP for jump and diffusion risk. 
We may refer $\widetilde{\mathbb{Q}}$ to the ``correct'' or ``true'' risk-neutral measure since the swap price is a $\widetilde{\mathbb{Q}}$-martingale without any approximations. In contrast, the ``classical'' or artificial risk-neutral measure,  $\mathbb{Q}$,  results from an approximation of the swap price leading in general not to the ``true'' pricing measure.
Consequently, any pricing methodology based on approximated averaging can easily be turned to the ``correct'' risk-neutral measure by an application of our MPDP.\\

We compare swap prices resulting from geometric averaging with swaps based on approximated averaging in line with
 with \cite{Kemper2022} and \cite{Bjerksund}.
We find that different averaging techniques lead to a pricing spread that stays untouched by measure changes.
In particular, the swap price based on geometric averaging turns out to be  smaller than the one resulting from approximated  averaging.
The spread itself can be characterized by a change of measure based on the MPDP as introduced by \cite{Kemper2022}.
As the MPDP leads to the true pricing measure, $\widetilde{\mathbb{Q}}$, the spread remediates  the approximated swap price and adjusts it downwards to the correct price of the swap contract.\\

We finally investigate the model under the physical measure.
To this end, we consider two types of models characterized, on the one hand, by seasonality in the delivery time (see \cite{fanelli2019seasonality}) and, on the other hand, by the Samuelson effect (see \cite{Samuelson}).
We adapt them to a jump setting, and provide the corresponding discretized swap price models.
Seasonal delivery dependence causes a MPDP that is constant over trading time and seasonal in delivery time.
In contrast, term-structure dependence analytically induces a decreasing behavior of the MPDP over trading time.
Hence, the closer we reach the expiration date, the more pronounced the MPDP, and the larger the pricing spread.
Consequently, the MPDP reduces risk caused by approximated averaging especially when the end of the maturity approaches.\\


To conclude, we expand the MPDP to the jump setting and investigate the MPDP influenced by typical characteristics of the electricity market.
We expect that a higher market share of renewables cause higher delivery-dependent seasonalities in the volatility and consequently leading to a growing MPDP.
This applies especially for Germany, having ambitious plans for future investments in renewable energy.
This increases the importance of the MPDP of diffusion risk, which has to be taken into account to ensure an accurate pricing procedure. However, this is a questions for future research.

\section*{Disclosure Statement}
No potential competing interest was reported by the authors.


\newpage
\appendix
\section*{\appendixname} \label{sec:appendix}
\addcontentsline{toc}{section}{\nameref{sec:appendix}}
\numberwithin{equation}{subsection}
\setcounter{equation}{0}
\renewcommand*{\thesubsection}{\Alph{subsection}} 
\subsection{Technical Requirements} \label{app:tech_requ3}
\begin{AssumptionAppendix} \label{ass:TechnicalRequirementsQ1}
For the model \eqref{eq:futuresQsol1}, we make the following assumptions to apply Itô's formula (see \cite{OksendalSulem}, cf.\ Theorem 1.16):
\begin{itemize}
	\item[$(i)$] For \mbox{$\mathcal{A}:=\{(t,\tau)\in~[0,\tau_2]^2\colon t\leq~\tau\}$} the functions $\sigma\colon \mathcal{A}\to \mathbb{R}^+$ and  $\eta\colon \mathcal{A}\to \mathbb{R}$ are  adapted such that the integrals exist, meaning that $\mathbb{Q}[\int_{0}^{t}\sigma^2(s,\tau)+ \int_{\mathbb{R}}\vert(e^{\eta(s,\tau)z}-1)\vert\ell^{\mathbb{Q}}(dz) ds<\infty ]=1$ for all $0\leq t\leq \tau$.
\end{itemize}
In order to ensure existence and uniqueness of solutions to  \Eqref{eq:futuresQ} (see \cite{OksendalSulem}, cf.\ Theorem 1.19), we further assume:
\begin{itemize}
\item[$(ii)$] (At most linear growth) There exists a constant $C_1<\infty$ such that
\begin{align}
\vert \sigma(t,\tau)x\vert^2 + \int_{\mathbb{R}} \vert (e^{\eta(t,\tau)z}-1)x\vert^2 \ell^{\mathbb{Q}}(dz) \leq C_1 (1+\vert x \vert^2)\;, \quad \forall x\in \mathbb{R}\;.
\end{align}
\item[$(iii)$] (Lipschitz continuity) There exists a constant $C_2<\infty$ such that
\begin{align}
\vert \sigma(t,\tau)x-\sigma(t,\tau)y\vert^2 + \int_{\mathbb{R}} \vert (e^{\eta(t,\tau)z}-1)x-(e^{\eta(t,\tau)z}-1)y\vert^2 \ell^{\mathbb{Q}}(dz) \leq C_2 (\vert x-y \vert^2)\;, \quad \forall x,y\in \mathbb{R}\;.
\end{align}
\end{itemize}
Hence, by \cite{OksendalSulem} (cf.\ Theorem 1.19), it follows that $\mathbb{E}_\mathbb{Q}[\vert f(t,\tau)\vert^2]<\infty$ for all $t\in[0,\tau]$.
By the Itô-Lévy Isometry (see \cite{OksendalSulem}, cf.\ Theorem 1.17) part $(iii)$ implies that 
\begin{align}
\mathbb{E}_\mathbb{Q}[f^2(t,\tau)] = \mathbb{E}_\mathbb{Q}\left[\int_{0}^{t}\sigma^2(v,\tau)f^2(v,\tau) +f^2(v,\tau) \int_{\mathbb{R}}(e^{\eta(v,\tau)z}-1)^2\ell^{\mathbb{Q}}(dz)dv \right]<\infty\;,
\end{align}
so that the square-integrability conditions are satisfied implying that  $f$ is a true martingale under $\mathbb{Q}$. 
\end{AssumptionAppendix}

\begin{AssumptionAppendix} \label{ass:TechnicalRequirementsQ2}
For the geometric weightening approach in \Eqref{eq:GeomAverage}, we need to apply the stochastic Fubini Theorem (see \cite{Protter2005}, cf.\ Theorem 65, Chapter IV. 6). Therefore, we assume that 
\begin{itemize}
\item[$(i)$] $\sigma(\cdot,\tau)$ and $\eta(\cdot,\tau)$ are $\mathcal{P}\times \mathcal{B}((\tau_1,\tau_2])$ measurable, where $\mathcal{P}$ is the predictable $\sigma$-algebra making all adapted, càglàd processes measurable,
\item[$(ii)$] $\mathbb{E}_{\mathbb{Q}}\left[\int_{0}^{\tau_1}\int_{\tau_1}^{\tau_2}\sigma^2(t,u)w(u,\tau_1,\tau_2)du~dt\right]<\infty$, 
\item[$(iii)$]$\mathbb{E}_{\mathbb{Q}}\left[\int_{0}^{\tau_1}\int_{\tau_1}^{\tau_2}\eta(t,u)w(u,\tau_1,\tau_2)du~dt\right]<\infty$, 
\item[$(iv)$] $\mathbb{E}_{\mathbb{Q}}\left[\int_{0}^{\tau_1}\int_{\mathbb{R}}\int_{\tau_1}^{\tau_2}(e^{\eta(t,u)z}-1)^2w(u,\tau_1,\tau_2)du~\ell^{\mathbb{Q}}(dz)dt\right]<\infty$,
\end{itemize}
such that the integrals still exist  and linear growth and Lipschitz continuity are satisfied (see \Cref{ass:TechnicalRequirementsQ1}).
\end{AssumptionAppendix}

\begin{AssumptionAppendix} \label{ass:TechnicalRequirementsQ3}
	To apply Girsanov's Theorem (see \cite{OksendalSulem}, cf.\ Theorem 1.35), we assume
	that $\Pi_1^{\mathbb{Q}\widetilde{\mathbb{Q}}}$ and $\Pi_2^{\mathbb{Q}\widetilde{\mathbb{Q}}}$ are  predictable, satisfying
\begin{itemize}
    \item[$(i)$]  $\mathbb{E}_{\mathbb{Q}}[\int_{0}^{\tau_1}\Pi_1^{\mathbb{Q}\widetilde{\mathbb{Q}}}(s,\tau_1,\tau_2)^2ds]<\infty$, such that $Z_1^{\mathbb{Q}\widetilde{\mathbb{Q}}}$ is a true martingale, and
    \item[$(ii)$]
     $\Pi_2^{\mathbb{Q}\widetilde{\mathbb{Q}}}(t,\tau_1,\tau_2)z\leq 1$ for all $t\in[0,\tau_1]$ and $\mathbb{E}_{\mathbb{Q}}[\int_{0}^{\tau_1}\ln(1-\Pi_2^{\mathbb{Q}\widetilde{\mathbb{Q}}}(s,\tau_1,\tau_2))+\Pi_2^{\mathbb{Q}\widetilde{\mathbb{Q}}}(s,\tau_1,\tau_2)ds]<\infty$, such that $Z_2^{\mathbb{Q}\widetilde{\mathbb{Q}}}$ is a true martingale.    
\end{itemize}
\end{AssumptionAppendix}
\newpage

\begin{AssumptionAppendix} \label{ass:TechnicalRequirementsP}
	For the model in \Eqref{eq:LogFuturesP}, we make the following assumptions to apply Itô's formula (see \cite{OksendalSulem}, cf.\ Theorem 1.16):
	\begin{itemize}
		\item[$(i)$] The functions $\mu\colon \mathcal{A}\to \mathbb{R}$, $\kappa\colon [0,\tau_1]\to \mathbb{R}^+$, $\sigma\colon \mathcal{A}\to \mathbb{R}^+$, and  $\eta\colon \mathcal{A}\to \mathbb{R}$ are  adapted such that the integrals exist, meaning that for all $0\leq t\leq \tau$, we have $\mathbb{P}[\int_{0}^{t} \mu^2(v,\tau)+  \kappa^2(v) +  \sigma^2(v,\tau)+  \eta^2(v,\tau)\int_{\mathbb{R}}z^2+e^{2\eta(v,\tau)z}\ell^{\mathbb{P}}(dz)  dv<\infty ]=1$.
	\end{itemize}
	In order to ensure existence and uniqueness of solutions to  \Eqref{eq:AdjOU2} (see \cite{OksendalSulem}, Theorem 1.19), we further assume:
	\begin{itemize}
		\item[$(ii)$] (At most linear growth) There exists a constant $C_1<\infty$ such that $ \forall x\in \mathbb{R}$:
		\begin{align}
		\vert  \mu(t,\tau)\vert^2+
		\vert \kappa(t)x\vert^2+
		\vert  \sigma(t,\tau)\vert^2 + \vert  \eta(t,\tau)\vert^2\int_{\mathbb{R}} \vert z\vert^2 \ell^{\mathbb{P}}(dz) +\int_{\mathbb{R}}\vert e^{ \eta(t,\tau)z}\vert^2 \ell^{\mathbb{P}}(dz) \leq C_1 (1+\vert x \vert^2)\;.
		\end{align}
		\item[$(iii)$] (Lipschitz continuity) There exists a constant $C_2<\infty$ such that
		\begin{align}
		\kappa^2(t)\vert x-y\vert^2 \leq C_2 (\vert x-y \vert^2)\;, \quad \forall x,y\in \mathbb{R}\;.
		\end{align}
	\end{itemize}
\end{AssumptionAppendix}


\begin{AssumptionAppendix} \label{ass:StochasticFubini}
For the geometric weightening approach in \Eqref{eq:GeomAverage} applied in \Cref{sec:Model}, we apply the stochastic Fubini Theorem (see \cite{Protter2005}, cf.\ Theorem 65, Chapter IV. 6). Therefore, we assume that 
\begin{itemize}
	\item[$(i)$]  $\kappa, \mu,\sigma,\eta$ are $\mathcal{P}\times \mathcal{B}((\tau_1,\tau_2])$ measurable, where $\mathcal{P}$ is the predictable $\sigma$-algebra making all adapted, càglàd processes measurable,
	\item[$(ii)$]
	$\mathbb{E}_{\mathbb{P}}\left[\int_{0}^{\tau_1}\int_{\tau_1}^{\tau_2} \mu(t,u)w(u,\tau_1,\tau_2)du~dt\right]<\infty$,
	\item[$(iii)$] $\mathbb{E}_{\mathbb{P}}\left[\int_{0}^{\tau_1}\int_{\tau_1}^{\tau_2} \sigma^2(t,u)w(u,\tau_1,\tau_2)du~dt\right]<\infty$,
	\item[$(iv)$] $\mathbb{E}_{\mathbb{P}}\left[\int_{0}^{\tau_1}\int_{\mathbb{R}}\int_{\tau_1}^{\tau_2}(e^{ \eta(t,u)z}-1)^2w(u,\tau_1,\tau_2)du~\ell^{\mathbb{P}}(dz)dt\right]<\infty$,
	\item[$(v)$]  $\mathbb{E}_{\mathbb{P}}\left[\int_{0}^{\tau_1}\int_{\tau_1}^{\tau_2} \eta(t,u)w(u,\tau_1,\tau_2)du~dt\right]<\infty$.
\end{itemize}
\end{AssumptionAppendix}

\begin{AssumptionAppendix}\label{ass:TechnicalRequirementsP3}
	To prove that $Z^{\mathbb{P}\widetilde{\mathbb{Q}}}$ is a true martingale, we assume that  $\kappa, \mu,\sigma,\eta$ are deterministic and that
	\begin{itemize}
	\item[$(i)$] $\Pi_2^{\mathbb{P}\widetilde{\mathbb{Q}}}(t,\tau_1,\tau_2)z<1$ for $\ell^{\mathbb{P}}$-a.e.\ $z\in\mathbb{R}$ and each $t\in[0,\tau_1]$,
	\item[$(ii)$] $\ell^{\mathbb{P}}$ has fourth moment, that
	is $\int_{\mathbb{R}}z^4\ell^{\mathbb{P}}(dz) < \infty$,
	\item[$(iii)$] $\int_{0}^{\tau_1}\int_{\tau_1}^{\tau_2}w(u,\tau_1,\tau_2)\mu^2(t,u) du~dt<\infty $,
	\item[$(iv)$] $\int_{0}^{\tau_1}\int_{\tau_1}^{\tau_2}w(u,\tau_1,\tau_2) \eta^2(t,u) du~dt<\infty $,
	\item[$(v)$] $\int_{0}^{\tau_1}\int_{\tau_1}^{\tau_2}w(u,\tau_1,\tau_2)\sigma^4(t,u) du~dt<\infty $,
	\item[$(vi)$] $\int_{0}^{\tau_1}\kappa^2(t)dt<\infty $.
	\end{itemize} 
\end{AssumptionAppendix} \newpage
\subsection{Proof of  \Cref{prop:ZisMartingale}}\label{app:Proof_MeasureChangeF}
Inspired by \cite{Benth2019}, we prove 
that $\mathbb{E}_{{\mathbb{P}}}[Z^{\mathbb{P}\widetilde{\mathbb{Q}}}(\tau_1,\tau_1,\tau_2)]=1$  and expand their Theorem 3.5 to a geometric setting with stochastic volatility in order to address settings as in \cite{Kemper2022}.
For the scope of the proof, we consider the swap price $F$ from  \Cref{lem:SwapPGeom} characterized by  stochastic volatility of the form
\begin{align}
\sigma (t,\tau) \sqrt{\nu (t)}\;,
\end{align} 
where $\sigma(t,\tau)$ is deterministic and $\nu$ is the stochastic volatility that is modeled as a Cox-Ingersoll-Ross process evolving as
\begin{align}
d\nu (t) = \kappa_\nu \left(\theta_\nu -\nu (t)\right)dt + \sigma_\nu \sqrt{\nu (t)}dB^\mathbb{P}_t\;, \label{eq:CIR}
\end{align}
for $\nu(0)=\nu_0>0$, where $B^\mathbb{P}$ and $\widetilde{J}^{\mathbb{P}}$ are independent of each other and $B^\mathbb{P}$ and $W^{\mathbb{P}}$ are correlated.
In particular, we assume a correlation structure $d\langle W^{\mathbb{P}},B^{\mathbb{P}} \rangle_t = \rho dt$ where $\rho \in(-1,1)$ such that we can rewrite $B^{\mathbb{P}}=\rho W ^{\mathbb{P}}+\sqrt{1-\rho^2}\bar{B}^{\mathbb{P}}$ for $\bar{B} ^{\mathbb{P}}\independent W^{\mathbb{P}}$. 
Moreover, we assume that $\kappa _\nu,\theta_\nu ,\sigma_\nu>0$ satisfy the extended Feller condition, i.e., $\sigma_\nu^2<\kappa _\nu \theta_\nu $, to ensure that $\mathbb{E}_{\mathbb{P}} [\nu ^{-2}(t)]$ is bounded on the entire trading time $t\in[0,\tau_1]$ (see \cite{Dereich}, cf.\ Chapter 3).
Note, that the extended Feller implies the classical Feller condition (see \cite{KaratzasShreve1991}, cf.\ Chapter 5) ensuring that the volatility stays positive.\\

We proceed in the following steps:
\begin{enumerate}
\item Derivation of a new risk-neutral measure $\widetilde{\mathbb{Q}}^n$ through a stopping time $\hat{\tau}_n$.
\item Proof that $\mathbb{E}_{\widetilde{\mathbb{Q}}}[Z^{\mathbb{P}\widetilde{\mathbb{Q}}}(\tau_1,\tau_1,\tau_2)]$ is lower boundend, i.e., $$\mathbb{E}_{\widetilde{\mathbb{Q}}}\left[Z^{\mathbb{P}\widetilde{\mathbb{Q}}}(\tau_1,\tau_1,\tau_2)\right]\geq
1-\frac1n \mathbb{E}_{\widetilde{\mathbb{Q}}^n}\left[\sup_s\ln F (s,\tau_1,\tau_2)\right]-\frac1n \mathbb{E}_{\widetilde{\mathbb{Q}}^n}\left[\sup_s\nu ^{-1}(s)\right]-\frac1n \mathbb{E}_{\widetilde{\mathbb{Q}}^n}\left[\sup_s\nu (s)\right]\;.$$
\item Proof that there exist upper boundaries for $\mathbb{E}_{\widetilde{\mathbb{Q}}^n}[\sup_s\ln F (s,\tau_1,\tau_2)]$, $\mathbb{E}_{\widetilde{\mathbb{Q}}^n}[\sup_s\nu ^{-1}(s)]$, and $\mathbb{E}_{\widetilde{\mathbb{Q}}^n}[\sup_s\nu (s)]$,  that are independent of $n$.\\
\end{enumerate}

\textbf{1. Derivation of $\widetilde{\mathbb{Q}}^n$.}
Similar to \cite{Benth2019}, we set $g(z):=(1+z)\log(1+z)-z$ and define the predictable compensator of 
$\frac12\langle H^c,H^c \rangle + \sum_{t\leq \cdot} g(\Delta H(t))$ by
\begin{align*}
C(t,\tau_1,\tau_2):=\frac12 \int_{0}^{t}  \pi_1 ^{\mathbb{P}\widetilde{\mathbb{Q}}}(s,\tau_1,\tau_2)^2 + \pi_\nu ^{\mathbb{P}\widetilde{\mathbb{Q}}}(s,\tau_1,\tau_2)^2
ds + \int_{0}^{t}\int_{\mathbb{R}}g(\pi_2^{\mathbb{P}\widetilde{\mathbb{Q}}}(s,\tau_1,\tau_2)z)\ell^{{\mathbb{P}}}(dz)ds\;, 
\end{align*}
where $H$ from \Eqref{eq:H} now embraces stochastic volatility such that  
\begin{align*}
H(t,\tau_1,\tau_2):=\int_{0}^{t} \pi_1 ^{\mathbb{P}\widetilde{\mathbb{Q}}}(s,\tau_1,\tau_2)dW ^{\mathbb{P}}_s+\int_{0}^{t}\int_{\mathbb{R}} \pi_2^{\mathbb{P}\widetilde{\mathbb{Q}}}(s,\tau_1,\tau_2)z\widetilde{N}^{\mathbb{P}}(ds,dz)+ \int_{0}^{t} \pi_\nu ^{\mathbb{P}\widetilde{\mathbb{Q}}}(s,\tau_1,\tau_2)d\bar{B} ^{\mathbb{P}}_s\;.
\end{align*}
Note that this stochastic volatility setting covers a three-dimensional market price of risk $\pi:=(\pi_1,\pi_2,\pi_\nu)$ for all independent random parts $W^{\mathbb{P}},\widetilde{J}^{\mathbb{P}},\bar{B}^{\mathbb{P}}$.
As we are in an incomplete setting, we choose the market price of volatility risk, $\pi_\nu$, such that the market price of risk admits the same structure as in the Heston model, i.e., $\rho  \pi_1^{\mathbb{P}\widetilde{\mathbb{Q}}} +\sqrt{1-\rho^2}\pi_\nu^{\mathbb{P}\widetilde{\mathbb{Q}}}=\frac{\delta_\nu}{\sigma_\nu }\sqrt{\nu (t)}$ (see \cite{Heston}). Now let us define a sequence of stopping times
\begin{align}
\hat{\tau}_n:=\inf \big\{ t\in[0,\tau_1]: \vert \ln F(t,\tau_1,\tau_2)\vert\geq n, \text{ or } \vert \nu^{-1}(t)\vert \geq n,  \text{ or } \vert \nu(t)\vert \geq n \big\}\;, \label{eq:StoppingTime}
\end{align}
and observe that for every $n\in\mathbb{N}$, the stopped process $C(t\wedge\hat{\tau}_n,\tau_1,\tau_2)$ is bounded.
Hence, by \cite{LepingleMEmin} (cf.\ Theorem III.1), we know that $Z^{\mathbb{P}\widetilde{\mathbb{Q}}}(t\wedge\hat{\tau}_n,\tau_1,\tau_2)$ is a uniformly integrable martingale such that we can define the probability measure $\widetilde{\mathbb{Q}}^n$ by
\begin{align}
\frac{d\widetilde{\mathbb{Q}}^n}{d\mathbb{P}} := Z^{\mathbb{P}\widetilde{\mathbb{Q}}}(\tau_1\wedge\hat{\tau}_n,\tau_1,\tau_2)\;.\label{eq:MeasureChangeQn}
\end{align}

\textbf{2. Proof of lower boundary of $\mathbb{E}_{{\mathbb{P}}}[Z^{\mathbb{P}\widetilde{\mathbb{Q}}}(\tau_1)]$.} 
First, $Z^{\mathbb{P}\widetilde{\mathbb{Q}}}$ is a positive local martingale by the assumption that $\pi_2^{\mathbb{P}\widetilde{\mathbb{Q}}}(t,\tau_1,\tau_2)\geq -1$ for all $t\in[0,\tau_1]$. Hence, it is a supermartingale, so that
we know the upper boundary   for $\tau_1\geq 0$:
\begin{align*}
\mathbb{E}_{\mathbb{P}}[Z^{\mathbb{P}\widetilde{\mathbb{Q}}}(\tau_1,\tau_1,\tau_2)] \leq \mathbb{E}_{\mathbb{P}}[Z^{\mathbb{P}\widetilde{\mathbb{Q}}}(0,\tau_1,\tau_2)] =1\;.
\end{align*}
Next, we consider the lower boundary, following \cite{Benth2019}:
\begin{align*}
\mathbb{E}_{\mathbb{P}}[Z^{\mathbb{P}\widetilde{\mathbb{Q}}}(\tau_1,\tau_1,\tau_2)] \geq
\mathbb{E}_{\mathbb{P}}[Z^{\mathbb{P}\widetilde{\mathbb{Q}}}(\tau_1,\tau_1,\tau_2) \mathbbm{1}_{\hat{\tau}_n > \tau_1}] 
=\mathbb{E}_{\mathbb{P}}[Z^{\mathbb{P}\widetilde{\mathbb{Q}}}(\tau_1\wedge \hat{\tau}_n,\tau_1,\tau_2) \mathbbm{1}_{\hat{\tau}_n > \tau_1}]= \widetilde{\mathbb{Q}}^n[\hat{\tau}_n > \tau_1]\;,
\end{align*}
where the last equality follows from the change of measure defined in Step 1 (see \Eqref{eq:MeasureChangeQn}). By definition of the stopping time $\hat{\tau}_n$ (see \Eqref{eq:StoppingTime}), we deduce
\begin{align*}
\mathbb{E}_{\mathbb{P}}[Z^{\mathbb{P}\widetilde{\mathbb{Q}}}(\tau_1,\tau_1,\tau_2)] \geq & 1- \widetilde{\mathbb{Q}}^n[\hat{\tau}_n \leq \tau_1]\\ \geq & 1- \left(\widetilde{\mathbb{Q}}^n\left[\sup_{s\in[0,\tau_1]}\ln F (s,\tau_1,\tau_2)\geq n\right]+\widetilde{\mathbb{Q}}^n\left[\sup_{s\in[0,\tau_1]}\nu ^{-1}(s)\geq n\right] +\widetilde{\mathbb{Q}}^n\left[\sup_{s\in[0,\tau_1]}\nu (s)\geq n\right] \right)\\
\geq & 1- \frac1n  \left(\mathbb{E}_{\widetilde{\mathbb{Q}}^n}\left[\sup_{s\in[0,\tau_1]}\ln F (s,\tau_1,\tau_2)\right]+\mathbb{E}_{\widetilde{\mathbb{Q}}^n}\left[\sup_{s\in[0,\tau_1]}\nu ^{-1}(s)\right]+\mathbb{E}_{\widetilde{\mathbb{Q}}^n}\left[\sup_{s\in[0,\tau_1]}\nu (s)\right] \right)\;,
\end{align*}
where the last inequality follows from Markov's inequality.
If we show that the expectations on the right hand side have upper boundaries that are independent of $n\in\mathbb{N}$, then  $\mathbb{E}_{\mathbb{P}}[Z^{\mathbb{P}\widetilde{\mathbb{Q}}}(\tau_1,\tau_1,\tau_2)]=1$, which is addressed in the third step.\\

\textbf{3. Proof of upper boundaries.}
In order to identify upper boundaries under the measure $\widetilde{\mathbb{Q}}^n$ defined in \Eqref{eq:MeasureChangeQn}, we need to derive the dynamics of $\ln F$, $\nu ^{-1}$, and $\nu$  under $\widetilde{\mathbb{Q}}^n$. 
We apply Girsanov's theorem (see \cite{OksendalSulem}, cf.\ Theorem 1.35) to Equations~\eqref{eq:Y2tilde} and \eqref{eq:CIR}, where 
\begin{align*}
W^{\widetilde{\mathbb{Q}}^n}_t &= W^{\mathbb{P}}_t + \int_{0}^{t} \Pi_1 ^{\mathbb{P}\widetilde{\mathbb{Q}}}(s,\tau_1,\tau_2)\mathbbm{1}_{[0,\hat{\tau}_n]}(s)ds\;,\\
B^{\widetilde{\mathbb{Q}}^n}_t &= B^{\mathbb{P}}_t + \int_{0}^{t} \frac{\delta_\nu }{\sigma_\nu }\sqrt{\nu (s)}\mathbbm{1}_{[0,\hat{\tau}_n]}(s)ds\;,
\end{align*}
are correlated standard Brownian motions under $\widetilde{\mathbb{Q}}^n$ and 
\begin{align*}
\widetilde{N}^{\widetilde{\mathbb{Q}}^n}(dt,dz)=\widetilde{N}^{\mathbb{P}}(dt,dz) + \Pi_2^{\mathbb{P}\widetilde{\mathbb{Q}}}(t,\tau_1,\tau_2)\mathbbm{1}_{[0,\hat{\tau}_n]}(t)\ell^{\mathbb{P}}(dz)dt\;,
\end{align*}
is the $\widetilde{\mathbb{Q}}^n$-compensated Poisson random measure.
Moreover, by Ito's formula, we find 
\begin{align*}
d\nu ^{-1}(t) = \nu ^{-1}(t)\left( \kappa_\nu+\delta_\nu \mathbbm{1}_{[0,\hat{\tau}_n]}(t)-\nu ^{-1}(t)(\kappa_\nu \theta_\nu -\sigma_\nu^2)\right)dt 
- \sigma_\nu \nu ^{-\frac32}(t)dB ^{\widetilde{\mathbb{Q}}^n}_t\;.
\end{align*}
Hence, we can show 
\begingroup
\allowdisplaybreaks
\begin{align*}
&~\mathbb{E}_{\widetilde{\mathbb{Q}}^n}\left[\sup_{s\in[0,\tau_1]}\vert \nu ^{-1}(s) \vert\right]\\ 
\overset{(\star)}{\leq}&~ 
\frac{1}{\nu_0} + \mathbb{E}_{\widetilde{\mathbb{Q}}^n}\left[\sup_{s\in[0,\tau_1]}\int_{0}^{s}\nu ^{-1}(t)\left( \kappa_\nu+\delta_\nu \mathbbm{1}_{[0,\hat{\tau}_n]}(t)-\nu ^{-1}(t)(\kappa_\nu \theta_\nu -\sigma_\nu^2)\right)dt \right]+\mathbb{E}_{\widetilde{\mathbb{Q}}^n}\left[\sup_{s\in[0,\tau_1]}\int_{0}^{s}\sigma_\nu\nu ^{-\frac32}(t) dB ^{\widetilde{\mathbb{Q}}^n}_t\right]\\
\overset{(\star\star)}{\leq}&~ 
\frac{1}{\nu_0} + 
(\kappa+\No{\delta_\nu})\mathbb{E}_{\widetilde{\mathbb{Q}}^n}\left[\int_{0}^{\tau_1}\nu ^{-1}(t)dt\right]+(\kappa_\nu\theta_\nu -\sigma_\nu^2)\mathbb{E}_{\widetilde{\mathbb{Q}}^n}\left[\int_{0}^{\tau_1}\nu ^{-2}(t)dt\right]+\sigma_\nu n^{-\frac32}\mathbb{E}_{\widetilde{\mathbb{Q}}^n}\left[\sup_{s\in[0,\tau_1]}\int_{0}^{s}dB ^{\widetilde{\mathbb{Q}}^n}_t\right]\\
\overset{(\star\star\star)}{=}&~
\frac{1}{\nu_0} + 
(\kappa+\No{\delta_\nu})\int_{0}^{\tau_1}\mathbb{E}_{\widetilde{\mathbb{Q}}^n}\left[\nu ^{-1}(t)\right]dt+(\kappa_\nu\theta_\nu -\sigma_\nu^2)\int_{0}^{\tau_1}\mathbb{E}_{\widetilde{\mathbb{Q}}^n}\left[\nu ^{-2}(t)\right]dt\;.
\end{align*}
\endgroup
Inequality~$(\star)$ follows from the integral representation of $\nu ^{-1}$ and the triangle inequality. Inequality~$(\star\star)$ results from the fact, that the extended Feller condition is satisfied (i.e., $\sigma_\nu^2<\kappa_\nu\theta_\nu $) and that  $\nu^{-1}\leq n$ under~$\widetilde{\mathbb{Q}}^n$. Since both processes $\nu ^{-1}$ and $\nu ^{-2}$ are positive, the supremum disappears in the first two cases and the upper boundary is used. 
Equality~$(\star\star\star)$ is reached by stochastic Fubini to the first two integrals and the last term disappears.
From \cite{Dereich} (cf.\ Chapter 3), we know that the expectations of the inverse and the inverse quadratic stochastic volatility, $\mathbb{E}_{\widetilde{\mathbb{Q}}^n}\left[\nu ^{-1}(t)\right]$ and $\mathbb{E}_{\widetilde{\mathbb{Q}}^n}\left[\nu ^{-2}(t)\right]$, can be characterized explicitly and are bounded independently of $n$, as long as the extended Feller condition $\sigma_\nu^2 < \kappa_\nu \theta_\nu $ is satisfied. Hence, $\mathbb{E}_{\widetilde{\mathbb{Q}}^n}\left[\sup_{s\in[0,\tau_1]}\vert \nu^{-1}(s) \vert\right]\leq c_1 \independent n$.

Moreover, we can show that $\vert \nu  \vert^2$ is uniformly integrable:
{\small
\begingroup
\allowdisplaybreaks
\begin{align*}
&~\mathbb{E}_{\widetilde{\mathbb{Q}}^n}\left[\sup_{s\in[0,\tau_1]} \vert \nu (s) \vert^2\right]
=
\mathbb{E}_{\widetilde{\mathbb{Q}}^n}\left[\sup_{s\in[0,\tau_1]}
\left(\nu_0+\int_{0}^{s} \kappa_\nu\theta_\nu  -(\kappa_\nu +\delta_\nu  \mathbbm{1}_{[0,\hat{\tau}_n]}(t))\nu (t)dt + \int_{0}^{s}\sigma_\nu \sqrt{\nu (t)}dB ^{\widetilde{\mathbb{Q}}^n}(t)\right)^2\right]\\
\overset{(\star)}{\leq}&~
4\Bigg( v_0^2+ 
\mathbb{E}_{\widetilde{\mathbb{Q}}^n}\left[\sup_{s\in[0,\tau_1]}
\left(\int_{0}^{s} \kappa_\nu\theta_\nu dt\right)^2
+\sup_{s\in[0,\tau_1]}
\left(\int_{0}^{s} (\kappa_\nu +\delta_\nu  \mathbbm{1}_{[0,\hat{\tau}_n]}(t))\nu (t)dt\right)^2
+\sup_{s\in[0,\tau_1]}
\left(\int_{0}^{s}\sigma_\nu \sqrt{\nu (t)}dB ^{\widetilde{\mathbb{Q}}^n}_t\right)^2\right]
\Bigg)\\
\overset{(\star\star)}{\leq}&~
4\Bigg( v_0^2+ 
4\mathbb{E}_{\widetilde{\mathbb{Q}}^n}\left[
\left(\int_{0}^{\tau_1} \kappa_\nu\theta_\nu dt\right)^2\right]
+4\mathbb{E}_{\widetilde{\mathbb{Q}}^n}\left[
\left(\int_{0}^{\tau_1} (\kappa_\nu +\delta_\nu  \mathbbm{1}_{[0,\hat{\tau}_n]}(t))\nu (t)dt\right)^2\right]
+4\mathbb{E}_{\widetilde{\mathbb{Q}}^n}\left[
\left(\int_{0}^{\tau_1}\sigma_\nu \sqrt{\nu (t)}dB ^{\widetilde{\mathbb{Q}}^n}_t\right)^2\right]
\Bigg)\\
\overset{(\star\star\star)}{\leq}&~
4\Bigg( v_0^2+ 
4\tau_1\int_{0}^{\tau_1} \kappa_\nu^2\theta_\nu^2dt
+4\tau_1 (\kappa_\nu +\No{\delta_\nu })^2\int_{0}^{\tau_1}\mathbb{E}_{\widetilde{\mathbb{Q}}^n}\left[ \sup_{s\in[0,t]} \nu (s)^2\right]dt
+4\sigma_\nu^2\mathbb{E}_{\widetilde{\mathbb{Q}}^n}\left[
\int_{0}^{\tau_1} \nu (t)dt\right]\;,
\Bigg)
\end{align*}
\endgroup
}
where the first equality represents the integral version of $\nu$. Inequality~$(\star)$ results from the Cauchy-Schwartz inequality to the sum
and an application of the triangle inequality. 
We apply Doob's inequality to all expectations in Inequality~$(\star\star)$. 
In Inequality~$(\star\star\star)$, we apply the Cauchy-Schwartz inequality to the first and second integral and apply Ito's isometry to the last summand.  
We finish with the stochastic Fubini to the  second integral while making the integrand even bigger. 
Note that for the last summand, we have $\mathbb{E}_{\widetilde{\mathbb{Q}}^n}\left[
\int_{0}^{\tau_1} \nu (t)dt\right]\leq \tilde{c}_{\nu}\independent n$ since we can find explicit expressions in \cite{cont_tankov2003financial} (cf.\ Chapter 15).
Setting $c_{\nu}:=4v_0^2+ 
16\tau_1^2 \kappa_\nu^2\theta_\nu ^2
+16\sigma_\nu^2\tilde{c}_{\nu}$, then, by Gronwall, we receive
$
\mathbb{E}_{\widetilde{\mathbb{Q}}^n}\left[\sup_{s\in[0,\tau_1]} \vert \nu (s) \vert^2\right] \leq c_{\nu} e^{16(\kappa_\nu +\No{\delta_\nu })^2\tau_1^2 }=:c_2 \independent n$.

Next, we show that $\vert \ln F \vert^2$ is uniformly integrable:
{\small
\begingroup
\allowdisplaybreaks
\begin{align*}
&~\mathbb{E}_{\widetilde{\mathbb{Q}}^n}\left[\sup_{s\in[0,\tau_1]} \vert \ln F(s,\tau_1,\tau_2) \vert^2\right]\\
=&~ 
\mathbb{E}_{\widetilde{\mathbb{Q}}^n}\Bigg[\sup_{s\in[0,\tau_1]} \Big(\ln F(0,\tau_1,\tau_2)+ \int_{0}^{s}  \left(1-\mathbbm{1}_{[0,\hat{\tau}_n]}(t)\right)\left(\mathbb{E} [ \mu(t,U)]- \kappa(t)\ln F(t,\tau_1,\tau_2)\right)dt
-\int_{0}^{s}\frac12\mathbb{E} [\sigma (t,U) ]^2\nu (t)\mathbbm{1}_{[0,\hat{\tau}_n]}(t)dt\\&~
+\int_{0}^{s}\mathbb{E} [\sigma (t,U) ]\sqrt{\nu (t)}dW^{\widetilde{\mathbb{Q}}^n}_t
+\int_{0}^{s}\mathbb{E} [ \eta(t,U)]d\widetilde{J}^{\widetilde{\mathbb{Q}}^n}_t\\&~
-\int_{0}^{s}\mathbb{E} [ \eta(t,U)] \left(1- \frac{\mathbb{E} [ \eta(t,U)]\int_{\mathbb{R}}z\ell^{\mathbb{P}}(dz)}{\int_{\mathbb{R}}
	e^{\mathbb{E} [ \eta(t,U)]z}-1
	\ell^{\mathbb{P}}(dz)}\right)\mathbbm{1}_{[0,\hat{\tau}_n]}(t)\int_{\mathbb{R}}z  \ell^{\mathbb{P}}(dz)dt
\Big)^2\Bigg]\\
\overset{(\star)}{\leq}&~
7\Bigg(\ln F(0,\tau_1,\tau_2)^2 + \mathbb{E}_{\widetilde{\mathbb{Q}}^n}\left[
\sup_{s\in[0,\tau_1]}  \left(\int_{0}^{s}\left(1-\mathbbm{1}_{[0,\hat{\tau}_n]}(t)\right)\mathbb{E} [ \mu(t,U)]dt\right)^2\right]
\\&~
+
\mathbb{E}_{\widetilde{\mathbb{Q}}^n}\left[
\sup_{s\in[0,\tau_1]} \left(\int_{0}^{s}
\frac12\mathbb{E} [\sigma (t,U) ]^2\nu(t)\mathbbm{1}_{[0,\hat{\tau}_n]}(t)dt\right)^2\right]
+
\mathbb{E}_{\widetilde{\mathbb{Q}}^n}\left[\sup_{s\in[0,\tau_1]}\left(\int_{0}^{s}\left(1-\mathbbm{1}_{[0,\hat{\tau}_n]}(t)\right) \kappa(t)\ln F(t,\tau_1,\tau_2)dt\right)^2\right]\\&~
+
\mathbb{E}_{\widetilde{\mathbb{Q}}^n}\left[\sup_{s\in[0,\tau_1]}\left(\int_{0}^{s}\mathbb{E} [\sigma (t,U) ]\sqrt{\nu(t)}dW^{\widetilde{\mathbb{Q}}^n}_t \right)^2\right]
+\mathbb{E}_{\widetilde{\mathbb{Q}}^n}\left[\sup_{s\in[0,\tau_1]}\left(\int_{0}^{s}\mathbb{E} [ \eta(t,U)]d\widetilde{J}^{\widetilde{\mathbb{Q}}^n}_t
\right)^2\right]\\&~
+\mathbb{E}_{\widetilde{\mathbb{Q}}^n}\left[\sup_{s\in[0,\tau_1]}\left(\int_{0}^{s}\mathbb{E} [ \eta(t,U)] \left(1- \frac{\mathbb{E} [ \eta(t,U)]\int_{\mathbb{R}}z\ell^{\mathbb{P}}(dz)}{\int_{\mathbb{R}}
	e^{\mathbb{E} [ \eta(t,U)]z}-1
	\ell^{\mathbb{P}}(dz)}\right)\mathbbm{1}_{[0,\hat{\tau}_n]}(t)\int_{\mathbb{R}}z  \ell^{\mathbb{P}}(dz)dt
\right)^2\right]
\Bigg)
\\
\overset{(\star\star)}{\leq}&~
7\Bigg(\ln F(0,\tau_1,\tau_2)^2 + 4\mathbb{E}_{\widetilde{\mathbb{Q}}^n}\left[
 \left(\int_{0}^{\tau_1}\left(1-\mathbbm{1}_{[0,\hat{\tau}_n]}(t)\right)\mathbb{E} [ \mu(t,U)]dt\right)^2\right]
\\&~
+
4\mathbb{E}_{\widetilde{\mathbb{Q}}^n}\left[
\left(\int_{0}^{\tau_1}
\frac12\mathbb{E} [\sigma (t,U) ]^2\nu(t)\mathbbm{1}_{[0,\hat{\tau}_n]}(t)dt\right)^2\right]
+
4\mathbb{E}_{\widetilde{\mathbb{Q}}^n}\left[\left(\int_{0}^{\tau_1}\left(1-\mathbbm{1}_{[0,\hat{\tau}_n]}(t)\right) \kappa(t)\ln F(t,\tau_1,\tau_2)dt\right)^2\right]\\&~
+
4\mathbb{E}_{\widetilde{\mathbb{Q}}^n}\left[\left(\int_{0}^{\tau_1}\mathbb{E} [\sigma (t,U) ]\sqrt{\nu(t)}dW^{\widetilde{\mathbb{Q}}^n}_t \right)^2\right]
+4\mathbb{E}_{\widetilde{\mathbb{Q}}^n}\left[\left(\int_{0}^{\tau_1}\mathbb{E} [ \eta(t,U)]d\widetilde{J}^{\widetilde{\mathbb{Q}}^n}_t
\right)^2\right]\\&~
+4\mathbb{E}_{\widetilde{\mathbb{Q}}^n}\left[\left(\int_{0}^{\tau_1}\mathbb{E} [ \eta(t,U)] \left(1- \frac{\mathbb{E} [ \eta(t,U)]\int_{\mathbb{R}}z\ell^{\mathbb{P}}(dz)}{\int_{\mathbb{R}}
	e^{\mathbb{E} [ \eta(t,U)]z}-1
	\ell^{\mathbb{P}}(dz)}\right)\mathbbm{1}_{[0,\hat{\tau}_n]}(t)\int_{\mathbb{R}}z  \ell^{\mathbb{P}}(dz)dt
\right)^2\right]
\Bigg)
\\
\overset{(\star\star\star)}{\leq}&~
7\Bigg(\ln F(0,\tau_1,\tau_2)^2 + 4\tau_1\int_{0}^{\tau_1}\mathbb{E} [ \mu(t,U)]^2dt
\\&~
+
4\int_{0}^{\tau_1}\mathbb{E} [\sigma (t,U) ]^4dt~\mathbb{E}_{\widetilde{\mathbb{Q}}^n}\left[ \int_{0}^{\tau_1}\nu(t)^2dt\right]
+
4\int_{0}^{\tau_1} \kappa(t)^2dt~\mathbb{E}_{\widetilde{\mathbb{Q}}^n}\left[\int_{0}^{\tau_1}\ln F(t,\tau_1,\tau_2)^2dt\right]\\&~
+
4\mathbb{E}_{\widetilde{\mathbb{Q}}^n}\left[\left(\int_{0}^{\tau_1}\mathbb{E} [\sigma (t,U) ]\sqrt{\nu(t)}dW^{\widetilde{\mathbb{Q}}^n}_t \right)^2\right]
+4\mathbb{E}_{\widetilde{\mathbb{Q}}^n}\left[\left(\int_{0}^{\tau_1}\mathbb{E} [ \eta(t,U)]d\widetilde{J}^{\widetilde{\mathbb{Q}}^n}_t
\right)^2\right]\\&~
+4\int_{0}^{\tau_1}\mathbb{E} [ \eta(t,U)]^2dt
\mathbb{E}_{\widetilde{\mathbb{Q}}^n}\left[\int_{0}^{\tau_1}\left(1-\mathbb{E} [ \eta(t,U)]^2 \frac{\left(\int_{\mathbb{R}}z\ell^{\mathbb{P}}(dz)\right)^2}{\left(\int_{\mathbb{R}}
	e^{\mathbb{E} [ \eta(t,U)]z}-1
	\ell^{\mathbb{P}}(dz)\right)^2}\right)\left(\int_{\mathbb{R}}z \ell^{\mathbb{P}}(dz)\right)^2dt\right]
\Bigg)
\\
\overset{}{\leq}&~
7\Bigg(\ln F(0,\tau_1,\tau_2)^2 + 4\tau_1\int_{0}^{\tau_1}\mathbb{E} [ \mu(t,U)]^2dt
+
4c_2 \tau_1 \int_{0}^{\tau_1}\mathbb{E} [\sigma (t,U) ]^4dt\\&~
+
4\int_{0}^{\tau_1} \kappa(t)^2dt~\mathbb{E}_{\widetilde{\mathbb{Q}}^n}\left[\int_{0}^{\tau_1}\sup_{s\in[0,t]}\ln F(s,\tau_1,\tau_2)^2dt\right]
+
4\sqrt{\int_{0}^{\tau_1}\mathbb{E} [\sigma (t,U) ]^4dt} \sqrt{\tau_1 c_2}
+4\int_{0}^{\tau_1}\mathbb{E} [ \eta(t,U)]^2\int_{\mathbb{R}}z^2\ell^{\widetilde{\mathbb{Q}}^n}(dz)dt\\&~
+4\int_{0}^{\tau_1}\mathbb{E} [ \eta(t,U)]^2dt\int_{0}^{\tau_1}\left(\left(\int_{\mathbb{R}}z \ell^{\mathbb{P}}(dz)\right)^2+\mathbb{E} [ \eta(t,U)]^2 \frac{\left(\int_{\mathbb{R}}z\ell^{\mathbb{P}}(dz)\right)^4}{\left(\int_{\mathbb{R}}
	e^{\mathbb{E} [ \eta(t,U)]z}-1
	\ell^{\mathbb{P}}(dz)\right)^2}\right)dt
\Bigg)\;,\\
=:&~c_Y +28\int_{0}^{\tau_1} \kappa(t)^2dt~\mathbb{E}_{\widetilde{\mathbb{Q}}^n}\left[\int_{0}^{\tau_1}\sup_{s\in[0,t]}\ln F(s,\tau_1,\tau_2)^2dt\right]\;.
\end{align*}
\endgroup
}
The first equality represents the integral version of $\ln F$. Inequality~$(\star)$ results from the Cauchy-Schwartz inequality to the sum
and an application of the triangle inequality. 
We apply Doob's inequality to all expectations in Inequality~$(\star\star)$. 
In Inequality~$(\star\star\star)$, we apply the Cauchy-Schwartz inequality to the first three integrals.  
We finish with Itô-Lévy Isometry (see \cite{OksendalSulem}, cf.\ Theorem 1.17) to the last summand and an application of the stochastic Fubini theorem  to the fourth summand (including $\ln F$) while making the integrand even bigger. 
By the previous considerations, we know that  $\mathbb{E}_{\widetilde{\mathbb{Q}}^n}\left[\int_{0}^{\tau_1}\mathbb{E} [\sigma (t,U) ]^2\nu(t)dt \right]\leq\sqrt{\int_{0}^{\tau_1}\mathbb{E} [\sigma (t,U) ]^4dt}\mathbb{E}_{\widetilde{\mathbb{Q}}^n}\left[\sqrt{\int_{0}^{\tau_1}\nu(t)^2dt}\right]\leq \sqrt{\int_{0}^{\tau_1}\mathbb{E} [\sigma (t,U) ]^4dt} \sqrt{\tau_1 c_2}$ is bounded independently of  $n$ and that
$\int_{0}^{\tau_1}\mathbb{E} [\sigma (t,U) ]^4dt~\mathbb{E}_{\widetilde{\mathbb{Q}}^n}\left[ \int_{0}^{\tau_1}\nu(t)^2dt\right] \leq c_2 \tau_1 \int_{0}^{\tau_1}\mathbb{E} [\sigma (t,U) ]^4dt$ is independent of $n$.
By the choice of $c_Y$, an application of Gronwall's inequality yields
$
\mathbb{E}_{\mathbb{P}^n}\left[\sup_{s\in[0,\tau_1]} \vert \ln F(s,\tau_1,\tau_2) \vert^2\right] \leq c_{Y} e^{28\int_{0}^{\tau_1} \kappa(t)^2dt } =:c_3\independent n\;,
$
such that we have shown, that $Z^{\mathbb{P}\widetilde{\mathbb{Q}}}$ is indeed a true martingale.
\newline \newpage

{\small
\begin{thebibliography}{99}






\bibitem[Benth et al.\abstand(2008a)]{Benth2008}
\textsc{Benth, F.E., Benth, J.S., and Koekebakker, S.} (2008). Stochastic Modelling of Electricity and Related Markets. Vol. 11. \textit{World Scientific Publishing Company}.


\bibitem[Benth et al.\,(2007)]{Benth2007}
\textsc{Benth, F.E., Kallsen, J. and Meyer-Brandis, T.} (2007). 
A Non-Gaussian Ornstein-Uhlenbeck Process for Electricity Spot Price Modeling and Derivatives Pricing.
\textit{Applied Mathematical Finance} \textbf{14}(2), pp.\  153-169. 


\bibitem[Benth et al.\abstand(2014)]{Benth2014}
\textsc{Benth, F.E., Klüppelberg, C., Müller, G., and Vos, L.} (2014). Futures Pricing in Electricity Markets Based on Stable CARMA Spot Models. \textit{Energy Economics}
\textbf{44}, pp.\  392-406.

\bibitem[Benth and Koekebakker\abstand(2008)]{BenthKoekebakker2008}
\textsc{Benth, F.E. and Koekebakker, S.} (2008). Stochastic Modeling of Financial Electricity Contracts. \textit{Energy Economics} \textbf{30}(3), pp.\  1116-1157.

\bibitem[Benth and Paraschiv\abstand(2016)]{BenthParaschiv}
\textsc{Benth, F.E. and Paraschiv, F.} (2016). A Structural Model for Electricity Forward Prices. \textit{Working Paper}.

\bibitem[Benth et al.\abstand(2019)]{Benth2019}
\textsc{Benth, F.E., Piccirilli, M., and Vargiolu, T.} (2019). Mean-Reverting Additive Energy Forward Curves in a Heath–Jarrow–Morton Framework. \textit{Mathematics and Financial Economics} \textbf{13}(4), pp.\  543–577.


\bibitem[Bjerksund et al.\abstand(2010)]{Bjerksund}
\textsc{Bjerksund, P., Rasmussen, H., and Stensland, G.} (2010). Valuation and Risk Management in the Norwegian Electricity Market. In: Bjørndal, E.,  Bjørndal, M., Pardalos, P. M., and Rönnqvist, M. (Editors) \textit{Energy, Natural Resources and Environmental Economics}, pp.\  167-185.

\bibitem[Borovkova and Permana\abstand(2006)]{BorovkovaPermana2006}
\textsc{Borovkova, S. and Permana, F.J.} (2006). Modelling Electricity Prices by the Potential
Jump-diffusion. In: Shiryaev, A.N., Grossinho, M.R., Oliveira, P.E., Esquível, M.L. 
(Editors), \textit{Stochastic Finance. Springer} pp.\  239-263.

\bibitem[Borovkova and Schmeck\abstand(2017)]{BorovkovaSchmeck2017}
\textsc{Borovkova, S. and Schmeck, M.D.} (2017). Electricity Price Modeling with
Stochastic Time Change. \textit{Energy Economics} \textbf{63}, pp.\  51-65.


\bibitem[Burger et al.\abstand(2004)]{BurgerMuller2004}
\textsc{Burger, M., Klar, B., Müller, A., and Schindlmayr, G.} (2004).
A Spot Market Model for Pricing Derivatives in Electricity Markets.
\textit{Quantitative Finance} \textbf{4}(1), pp.\  109-122.




\bibitem[Cartea and Figueroa\abstand(2005)]{CarteaFiguora2005}
\textsc{Cartea, A. and Figueroa, M.G.} (2005). Pricing in Electricity Markets: A Mean Reverting Jump Diffusion Model with Seasonality. \textit{Applied Mathematical Finance} \textbf{12}(4).

\bibitem[Cartea and Villaplana\abstand(2008)]{Cartea2008}
\textsc{Cartea, A. and Villaplana, P.} (2008. Spot Price Modeling and the Valuation of Electricity Forward Contracts: The Role of Demand and Capacity. \textit{Journal of Banking \& Finance} \textbf{32}(12), pp.\ 2502-2519.

\bibitem[Clewlow and Strickland\abstand(1999)]{ClewlowStrickland1999}
\textsc{Clewlow, L. and Strickland, C.} (1999). Valuing Energy Options in a One Factor Model fitted to Forward Prices. \textit{SSRN} \textbf{160608}.

\bibitem[Cont and Tankov\abstand(2004)]{cont_tankov2003financial}
\textsc{Cont, R. and Tankov, P.} (2004). Financial Modelling with Jump Processes. \textit{Chapman and Hall/CRC}. 


\bibitem[Cuchiero et al.\abstand(2022)]{CuchieroPersioGuidaSvalute2022}
\textsc{Cuchiero, C., Persio,  L.D.,  Guida, F., and Svaluto-Ferro, S.} (2022). Measure-valued Processes for Energy Markets. \textit{arXiv:2210.09331v1}.


\bibitem[Dereich et al.\abstand(2012)]{Dereich}
\textsc{Dereich, S., Neuenkirch, A., and Szpruch, L.} (2012). An Euler-type method for the strong approximation of the Cox–Ingersoll–Ross process. \textit{The Royal Society} \textbf{468}, pp.\  1105-1115.




\bibitem[Escribano et al.\abstand(2011)]{Escribano2011}
\textsc{Escribano, A., Peña, J.I., and Villaplana, P.} (2011). Modelling Electricity Prices: International Evidence.  \textit{Oxford Bulletin of Economics and Statistics} \textbf{73}(5), pp.\  622-650.

\bibitem[Fanelli and Schmeck\abstand(2019)]{fanelli2019seasonality}
\textsc{Fanelli, V. and Schmeck, M.D.} (2019). On the Seasonality in the Implied Volatility of Electricity Options.  \textit{Quantitative Finance} \textbf{19}(8), pp.\  1321-1337.





\bibitem[Gray and Pitts\abstand(2012)]{GrayPitts2012}
\textsc{Gray, R.J.  and Pitts, S.M.} (2012). Risk Modelling in General Insurance: From Principles to Practice. \textit{Cambridge University Press}.

\bibitem[Heath et al.\abstand(1990)]{HJM}
\textsc{Heath, D., Jarrow, R., and Morton, A.} (1990). Bond Pricing and the Term Structure of Interest Rates: A New Methodology for Contingent Claims Valuation. \textit{Econometrica} \textbf{60}(1), pp.\  77-105.

\bibitem[Heston\abstand(1993)]{Heston}
\textsc{Heston, S.L.} (1993). A Closed-Form Solution for Options with Stochastic Volatility with Applications to Bond and Currency Options. \textit{The Review of Financial Studies} \textbf{6}(2), pp.\  327-343.

\bibitem[Hinderks et al.\abstand(2020)]{Hinderks2020}
\textsc{Hinderks, W.J., Korn, R., and Wagner, A.} (2020).
A Structural Heath–Jarrow–Morton Framework for
consistent Intraday Spot and Futures Electricity
Prices. \textit{Quantitative Finance} \textbf{20}(3), pp.\   347-357.


\bibitem[Jaeck and Lautier\abstand(2016)]{JaeckLautier}
\textsc{Jaeck, E. and Lautier, D.} (2016). Volatility in Electricity Derivative Markets: The Samuelson
Effect Revisited. \textit{Energy Economics} \textbf{59}, pp.\  300-313.


\bibitem[Karatzas and Shreve\abstand(1991)]{KaratzasShreve1991}
\textsc{Karatzas, I. and Shreve, S.E.} (1991). Brownian Motion and Stochastic Calculus. \textit{Springer}; 2nd edition. 

\bibitem[Kemna and Vorst\abstand(1990)]{KemnaVorst1990}
\textsc{Kemna, A.G.Z. and Vorst, A.C.F.} (1990). A Pricing Method for Options based on Average Asset Values. \textit{Journal of Banking and Finance} \textbf{14}(1), pp.\  113-129.

\bibitem[Kemper et al.\abstand(2022)]{Kemper2022}
\textsc{Kemper, A., Schmeck, M.D., and Kh.Balci, A.} (2022). The Market Price of Risk for Delivery Periods: Pricing Swaps and Options in Electricity Markets.  \textit{Energy Economics} \textbf{113}(106221).


\bibitem[Kiesel et al.\abstand(2009)]{kiesel}
\textsc{Kiesel, R., Schindlmayr, G., and Börger, R.H.} (2009). A Two-Factor Model
for the Electricity Forward Market. \textit{Quantitative Finance} \textbf{9}(3),
279-287.

\bibitem[Kleisinger-Yu et al.\abstand(2020)]{KleisingerYu}
\textsc{Kleisinger-Yu, X., Komaric, V., Larsson, M., and Regez, M.} (2020). A Multifactor Polynomial Framework for Long-Term Electricity Forwards with Delivery Period.
\textit{SIAM Journal on Financial Mathematics} \textbf{11}(3), pp.\  928-957.

\bibitem[Klüppelberg et al.\abstand(2010)]{Klüppelberg2010}
\textsc{Klüppelberg, C., Meyer-Brandis, T. and Schmidt, A.} (2010). Electricity Spot Price Modelling with a View towards Extreme Spike Risk.
\textit{Quantitative Finance} \textbf{10}(9), pp.\  963–974.


\bibitem[Koekebakker and Ollmar\abstand(2005)]{KoekebakkerOllmar}
\textsc{Koekebakker, S. and Ollmar, F.} (2005). Forward Curve Dynamics in the Nordic Electricity
Market. \textit{Managerial Finance} \textbf{31}(6), pp.\  73-94.

\bibitem[Latini et al.\abstand(2019)]{Latini2019}
\textsc{Latini, L., Piccirilli, M., and Vargiolu, T.} (2019). Mean-reverting No-arbitrage Additive Models for Forward Curves in Energy Markets. \textit{Energy Economics} \textbf{79}, pp.\  157-170.

\bibitem[Lépingle and Mémin\abstand(1978)]{LepingleMEmin}
\textsc{Lépingle, D. and Mémin, J.} (1978). Sur l’intégrabilité Uniforme des Martingales Exponentielles. \textit{Zeitschrift für Wahrscheinlichkeitstheorie und verwandte Gebiete} \textbf{42}(3), pp.\  175–203.

\bibitem[Lucia and Schwartz\abstand(2002)]{LuciaSchwartz2002}
\textsc{Lucia, J.J. and Schwartz, E.} (2002). Electricity Prices and Power Derivatives: Evidence from the Nordic Power Exchange. \textit{Review of Derivatives
Research} \textbf{5}, pp.\  5–50.

\bibitem[Merton\abstand(1976)]{Merton1976}
\textsc{Merton, R.} (1976). Option Pricing when Underlying Stock Returns are Discontinuous. \textit{Journal of Financial
Economics} \textbf{3}(1-2), pp.\  125-144.




\bibitem[{\O}ksendal and Sulem\abstand(2007)]{OksendalSulem}
\textsc{{\O}ksendal, B., Sulem, A.} (2007). Applied Stochastic Control of Jump Diffusions. \textit{Springer}.

\bibitem[Papapantoleon\abstand(2008)]{Papapantoleon}
\textsc{Papapantoleon, A.} (2008). An Introduction to L\'{e}vy Processes with Applications in Finance. \textit{arXiv:0804.0482}.



\bibitem[Protter\abstand(2005)]{Protter2005}
\textsc{Protter, P.E.} (2005). Stochastic Differential Equations. In: Stochastic Integration and Differential Equations. Stochastic Modelling and Applied Probability, Vol. 21, \textit{Springer}.


\bibitem[Samuelson\abstand(1965)]{Samuelson}
\textsc{Samuelson, P.A.} (1965). Proof that Properly Anticipated Prices Fluctuate Randomly.  \textit{Industrial Management Review} \textbf{6}(2), pp.\  41–49.


\bibitem[Schneider and Tavin\abstand(2018)]{schneider2018samuelson}
\textsc{Schneider, L. and Tavin, B.} (2018). From the Samuelson Volatility Effect to a Samuelson Correlation Effect: An Analysis of Crude Oil Calendar Spread Options.  \textit{Journal of Banking and Finance} \textbf{95}, pp.\  185–202.

\bibitem[Shreve\abstand(2004)]{Shreve2004}
\textsc{Shreve, S.E.} (2004). Stochastic Calculus for Finance II. Continuous-Time Models. \textit{Springer Finance Series}. 




\end{thebibliography} 
}
\end{document}